\documentclass[referee]{aa}  

\usepackage{float}
\usepackage{graphicx}
\usepackage[varg]{txfonts}
\newcommand{\BLOS}{\ensuremath{B_\mathrm{LOS}}}
\newcommand{\lambdaB}{\ensuremath{\lambda_\mathrm{B}}}
\newcommand{\lambdaD}{\ensuremath{\lambda_\mathrm{D}}}

\begin{document}

   \title{Ubiquitous hundred-Gauss magnetic fields in solar spicules }


   \author{M. Kriginsky\inst{1,2} \and R. Oliver\inst{1,2} \and N. Freij\inst{1} \and D. Kuridze\inst{3} \and A. Asensio Ramos\inst{4} \and P. Antolin\inst{5} 
          }

   \institute{Departament de F\'\i sica, Universitat de les Illes Balears, E-07122 Palma de Mallorca, Spain
         \and
          Institute of Applied Computing \& Community Code (IAC3), UIB, Spain
          \and
            Department of Physics, Aberystwyth University, Ceredigion, SY23 3BZ, UK
          \and
            Instituto de Astrof\'\i sica de Canarias, La Laguna, Tenerife, Spain
           \and
             Department of Mathematics, Physics and Electrical Engineering, Northumbria University, Newcastle Upon Tyne, NE1 8ST, UK
                      }

   \date{Received ; accepted }

 
  \abstract
   {}
   {We use high-resolution spectropolarimetric observations in the \ion{Ca}{II} 8542 \AA\ line obtained with the Swedish 1-m Solar Telescope to study the magnetic field in solar spicules.}
   {The equations that result from the application of the Weak Field Approximation (WFA) to the radiative transfer equations are used to infer the line-of-sight component of the magnetic field ($B_{\mathrm{LOS}}$). Two restrictive conditions are imposed on the Stokes $I$ and~$V$ profiles at each pixel before they can be used in a Bayesian inversion to compute its $B_{\mathrm{LOS}}$.}
   {The line-of-sight (LOS) magnetic field component has been inferred in six data sets totalling 448 spectral scans in the \ion{Ca}{II} 8542~\AA\ line and containing both active region and quiet Sun areas, with values of hundreds of~G being abundantly inferred. There seems to be no difference, from the statistical point of view, between the magnetic field strength of spicules in the quiet Sun or near an active region. On the other hand, the \BLOS\ distributions present smaller values on the disk than off-limb, a fact that can be explained by the effect of superposition on the chromosphere of on-disk structures. We show that on-disk pixels in which the \BLOS\ is determined are possibly associated with spicular structures because these pixels are co-spatial with the magnetic field concentrations at the network boundaries and the sign of their \BLOS\ agrees with that of the underlying photosphere. We find that spicules in the vicinity of a sunspot have a magnetic field polarity (i.e. north or south) equal to that of the sunspot. This paper also contains an analysis of the effect of off-limb overlapping structures on the observed Stokes $I$ and $V$ parameters and the \BLOS\ obtained from the WFA. It is found that this value is equal to or smaller than the largest LOS magnetic field components of the two structures. In addition, using random \BLOS, Doppler velocities and line intensities of these two structures leads in $\simeq$50\% of the cases to Stokes $I$ and $V$ parameters unsuitable to be used with the WFA.}
   {Our results present a scarcity of LOS magnetic field components smaller than some 50~G, which must not be taken as evidence against the existence of these magnetic field strengths in spicules. This fact possibly arises as the consequence of signal superposition and noise in the data. We also suggest that the failure of previous works to infer the strong magnetic fields in spicules detected here is their coarser spatial and/or temporal resolution.}

   \keywords{Sun: chromosphere --
                Sun: magnetic fields --
                Weak field approximation
               }

   \titlerunning{Ubiquitous hundred-Gauss magnetic fields in solar spicules}
   \authorrunning{M. Kriginsky et al.}
   \maketitle
   
%

\section{Introduction}

Spicules are cold, dense, short-lived jet-like structures that shoot out from the chromosphere and reach coronal heights. Ever since their discovery by \citet{secchi}, they have remained a topic of great interest in solar physics and nowadays it is believed that they can play an important role mediating energy and mass transport to the corona \citep{Beckers1968,2011Sci...331...55D}. The term ``spicule'' was first coined by \citet{Beckers1968}, referring to objects visible off-limb, although they are present everywhere on the Sun and are ejected from the chromospheric network. Numerous reviews have been written as the knowledge about spicules has evolved \citep{Beckers1968,Sterling2000,2012RSPTA.370.3129R,Tsiropoula2012,HINODE2019}.

With the advance in instrumentation, more has been learnt about spicules. Of special importance was the introduction of the Hinode satellite \citep{2007SoPh..243....3K,2008SoPh..249..197S}, which provides high-cadence, high-resolution images necessary to properly resolve small structures. Perhaps the most important discovery about spicules in recent years is that they seem to come in two types \citep{2007adepontieu,pereira2012}. Type~I spicules are present mainly near active regions (ARs), show a rise and fall motion with constant acceleration and are visible in chromospheric lines during their whole lifetime. They are longer-lived and possess smaller mass flow velocities than their more energetic type II siblings, that are present everywhere on the Sun and are much more abundant. Type~II spicules seem to be accelerated upward and heated at the same time, hence they are observed at progressively hotter temperature lines, starting at chromospheric ones and then followed by transition region and coronal lines \citep{2009ApJ...705..272R,2007adepontieu,2017AGUFMSH43A2793D,2014ApJ...792L..15P,2016ApJ...820..124H}.

Direct magnetic field measurements of spicules are very scarce and have been done by using spectropolarimetric observations coupled with a theoretical modelling of the Zeeman and Hanle effects (see Tsiropoula et al. \citeyear{Tsiropoula2012} for a review and Socas-Navarro \& Elmore \citeyear{2005ApJ...619L.195S}, Centeno et al. \citeyear{Centeno2010} for a discussion of the magnetic field determination from spectropolarimetric data). In all works, which are described next, a spectrograph slit placed on off-limb spicules was used to acquire the data. \citet{TrujilloBueno2005} used spectropolarimetric observations in the \ion{He}{I} 10830 \AA\ triplet obtained with the Tenerife Infrared Polarimeter \citep[TIP;][]{1999ASPC..183..264M,2007ASPC..368..611C} at the German Vacuum Tower Telescope (VTT) in Observatorio del Teide with a net integration time of 209~s; this time corresponds to the averaging of 50 consecutive exposures. These authors performed both optically thin and optically thick theoretical modelings of the Hanle and Zeeman effects to infer magnetic field values of 10 G in a quiet Sun (QS) region. \citet{ariste} made spectropolarimetric observations in the \ion{He}{I} D$_3$ line taken with the Advanced Stokes Polarimeter \citep[ASP;][]{1992SPIE.1746...22E} at the Dunn Solar Telescope (DST) with the slit crossing an active region plage. The data have 6~s per exposure and a pixel scale of 0.29\arcsec~pixel$^{-1}$, although the seeing-induced image degradation yields a much worse spatial resolution \citep[see Fig.~6 of][]{ariste}. These authors reported the possibility of magnetic fields stronger than 30~G, even up to 40 G. \cite{2005ESASP.596E..82R,2006ASPC..358..448R} used a set of observations in the \ion{He}{I} D$_3$ line taken with the Zurich Imaging Polarimeter \citep[ZIMPOL;][]{2004A&A...422..703G} at the 45 cm Gregory-Coud\'e Telescope (GCT) at IRSOL with a 2-min cadence. They created a database of theoretical Stokes profiles using the quantum theory of the Hanle and Zeeman effects and upon inverting the observed spectropolarimetric signals, inferred magnetic field values of 50 G near an active region and 10 G in a quiet Sun area. \citet{Centeno2010} also used observations in the \ion{He}{I} 10830~\AA\ triplet at the VTT with time cadence and pixel scale of 56~s and 0.17\arcsec~pixel$^{-1}$, respectively; to increase the S/N, the data were spatially and temporally averaged to 1\arcsec and 45~min. Using the HAZEL code \citep{AsensioRamos2008} for the inversion, they could set a lower value of the magnetic field strength as high as 50 G in a quiet Sun region, namely the solar south pole. Further work on the spicules' magnetic field was carried out by \citet{orozco2015} with the help of spectropolarimetric \ion{He}{I} 10830~\AA\ data taken with the Tenerife Infrared Polarimeter II \citep[TIP-II][]{collados2007} at the VTT. The exposure time was about 10~s for a fixed slit position and the spatial resolution was estimated to be in the range 0.7--1\arcsec. A 9\arcsec\ wide area above the limb was scanned with the spectrograph slit, which allowed \citet{orozco2015} to infer the magnetic field vector with the HAZEL code and to study its variation with height. These authors concluded that the magnetic field is essentially vertical at the base of the chromosphere and changes its inclination to some 40$^\circ$ from the vertical direction at 2 Mm height. At the same time, its strength varies from 80~G to 30~G, on average, from the base of the chromosphere to a height of 3~Mm.

\begin{figure}
   \centering
   \includegraphics[width=8cm]{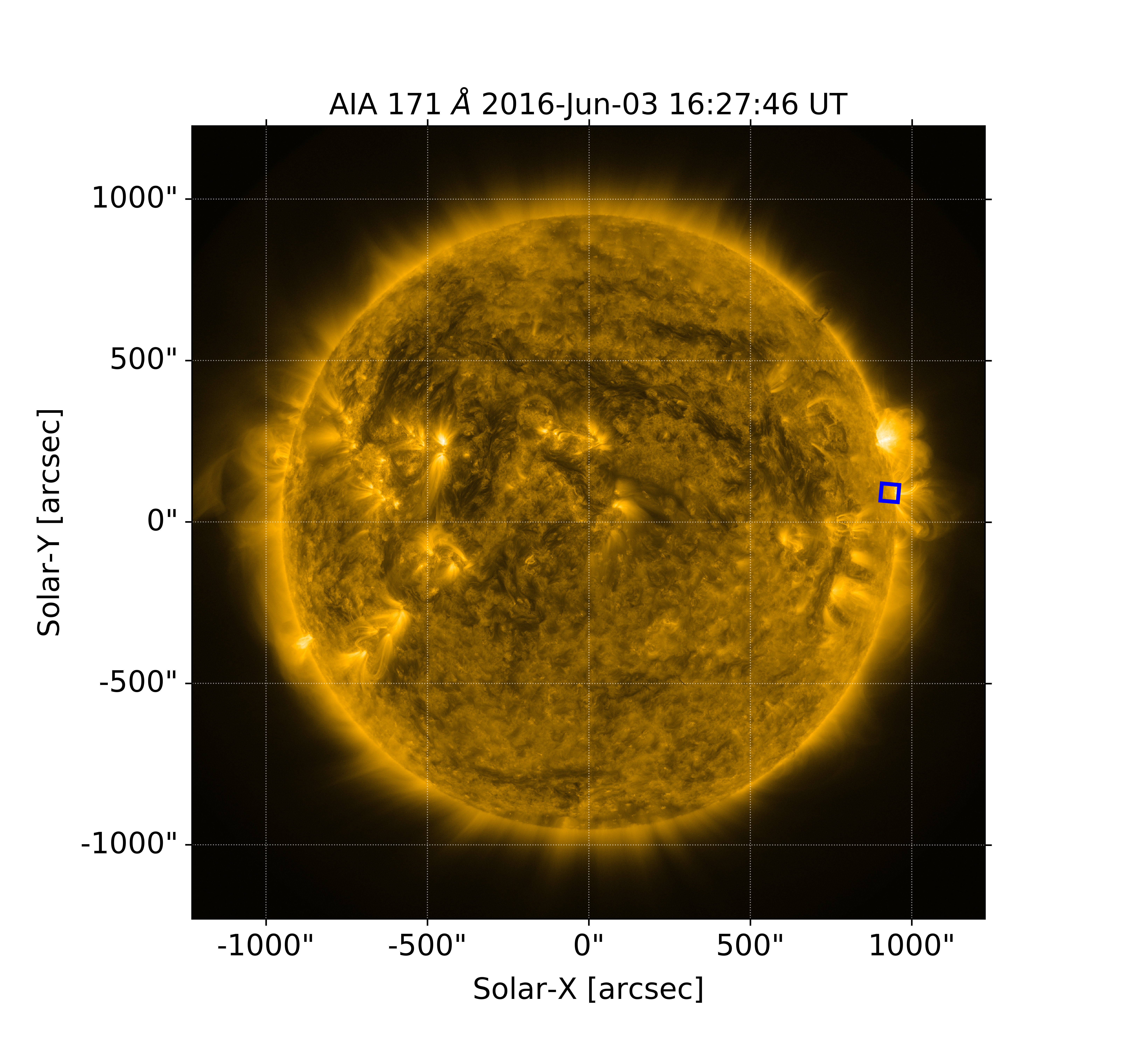}
   \includegraphics[width=8cm]{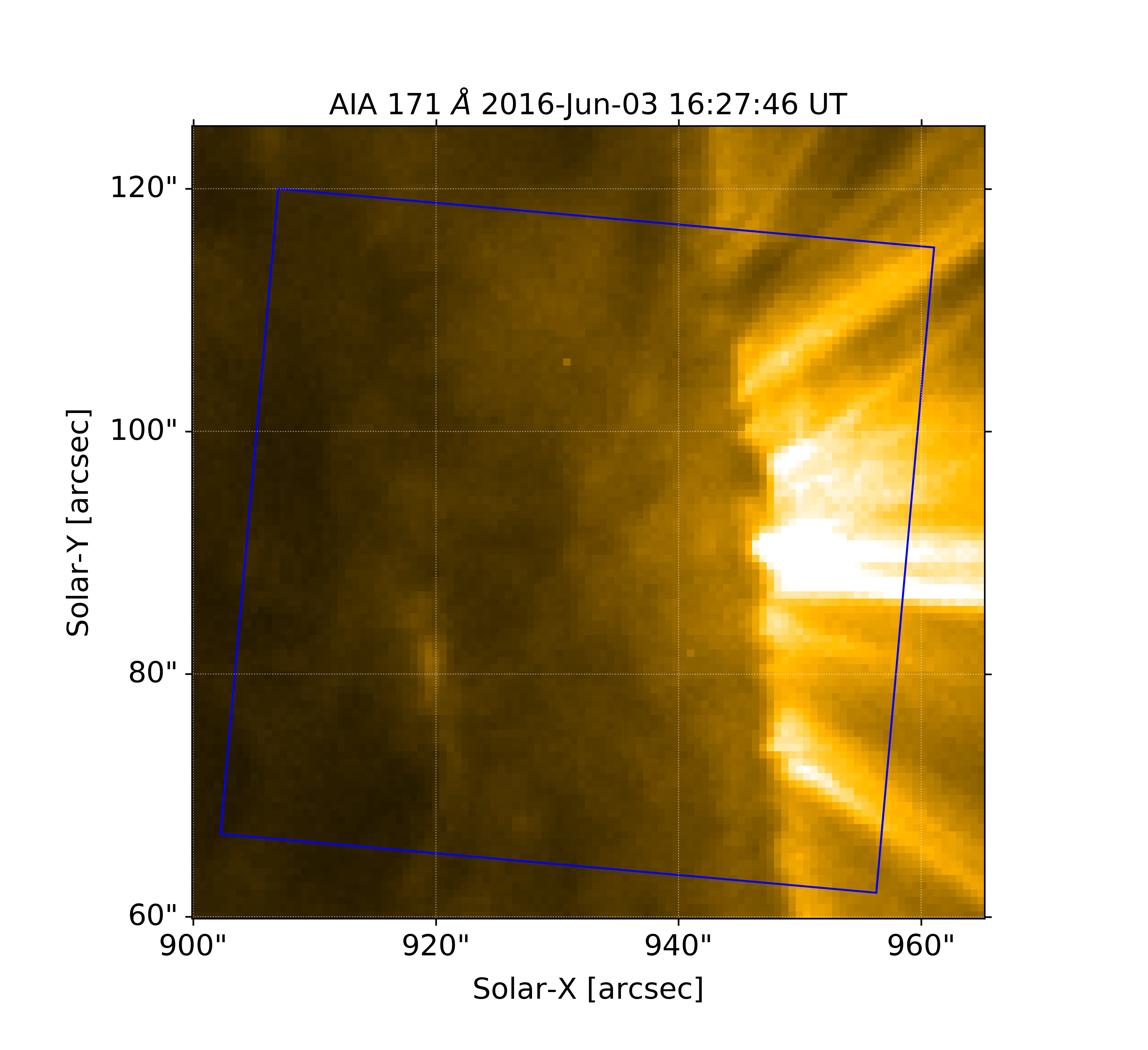}
   \includegraphics[width=8cm]{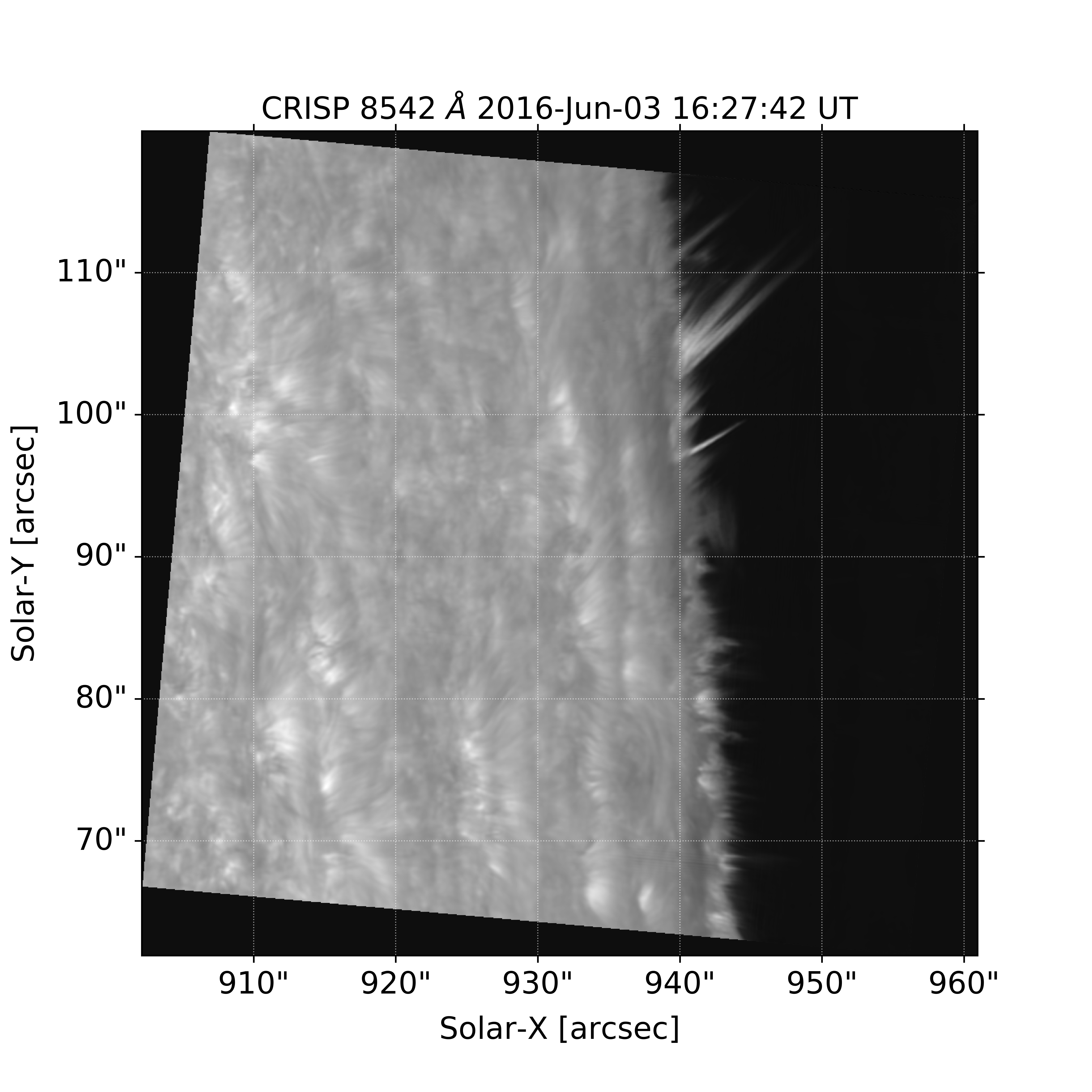}
   \caption{Context of the observations of 03 June 2016 at 16:27 UT. The top panel shows the CRISP field of view (blue square) on an SDO/AIA image in the 171 \AA\ filter. The middle panel shows a zoomed view of the top panel. The bottom panel shows a nearly co-temporal SST/CRISP image in the line centre of the \ion{Ca}{II} 8542 \AA\ line. Time is shown at the top of each frame and solar North is to the top.}
              \label{Figure:1}%
\end{figure}
  
\begin{figure*}
   \centering
   \includegraphics[width=14cm]{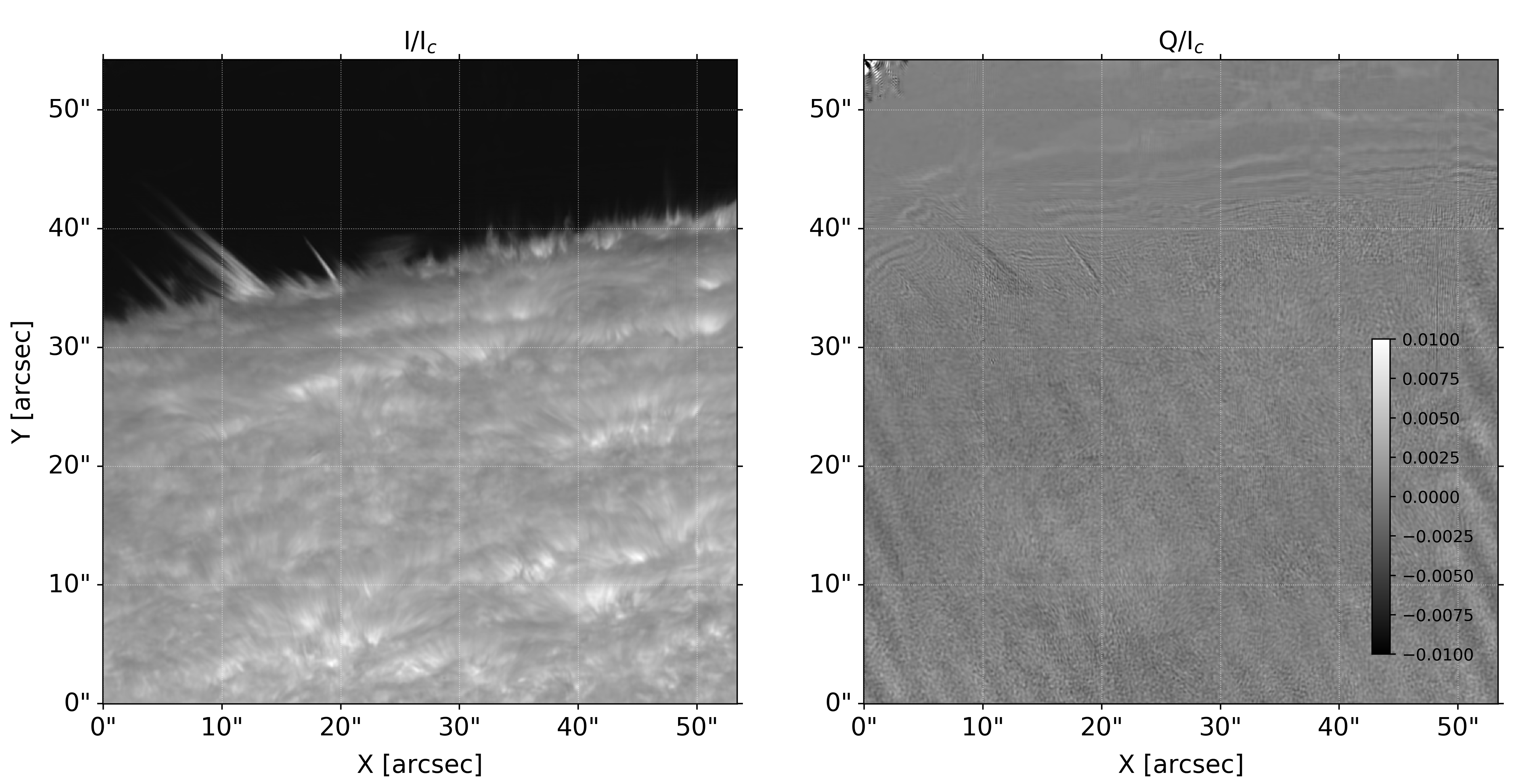}
   \includegraphics[width=14cm]{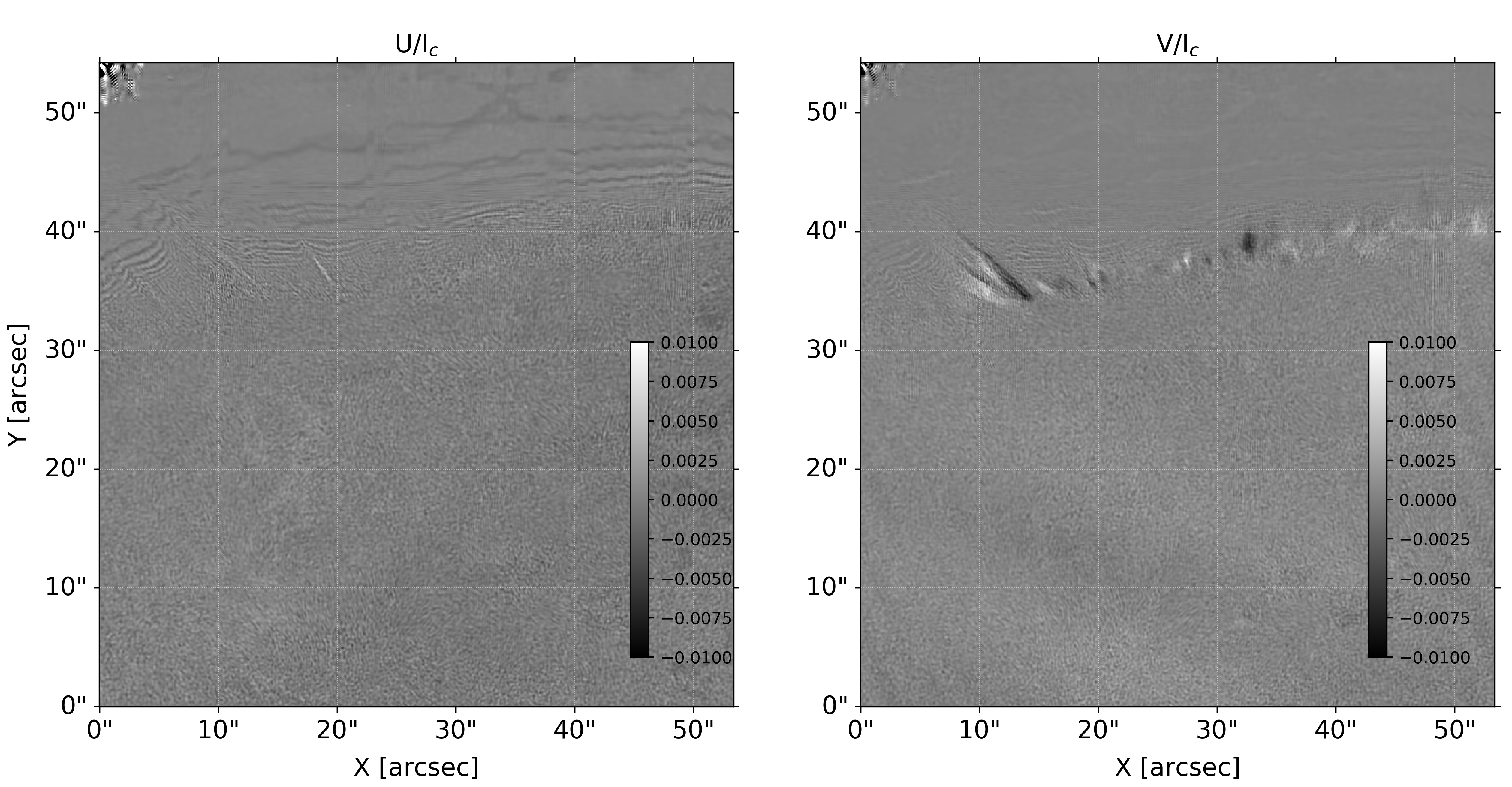}
   \caption{The Stokes parameters at line centre on 03 June 2016 at 16:27:42 UT. The parameters are normalised to the continuum intensity, $I_\mathrm{c}$, measured as the average of $I$ on the disk at the line wings ($|\Delta \lambda| > 1.25$~\AA). Solar North is to the left.}
              \label{Figure:2}%
\end{figure*}

Spicules are formed along chromospheric magnetic flux concentrations, therefore estimating their magnetic field is necessary in order to properly comprehend how such structures are formed and evolve. In this paper we analyse high-resolution imaging spectropolarimetric observations of solar spicules in the \ion{Ca}{II} 8542 \AA\ line, presented in Sect. \ref{sec:obs}. We use the Weak Field Approximation (WFA; Sect. \ref{sec:Data}), which is a computationally inexpensive method to calculate the magnetic field vector from the Stokes parameters when some conditions are met \citep[see, e.g.][]{landi,JennerholmHammar,delacruz2017,Centeno2018}. This technique has been used in the past with \ion{Ca}{II} 8542~\AA\ observations to infer the line-of-sight (LOS) magnetic field component (\BLOS) in a quiet Sun region near a decaying active region \citep{2007ApJ...663.1386P}, an active region filament \citep{2012SoPh..280...69H}, flares \citep{2012SoPh..280...69H,kuridze2018}, a sunspot umbra \citep{2013A&A...556A.115D} and flare coronal loops \citep{kuridze2019}. Our spectropolarimetric data contain Stokes $I$ and $V$ profiles above the noise level, which allows us to determine the LOS magnetic field component (Sect.~\ref{results}). The computed \BLOS\ values are of the order of hundreds of~G, i.e. substantially higher than previous estimates for solar spicules. In the discussion, Sect.~\ref{sec:discussion}, we use a simple model to assess the effect on the Stokes $I$ and $V$ parameters of two structures overlapping along the line-of-sight and the \BLOS\ derived from the WFA. Furthermore, we show that the measured LOS magnetic fields correspond to spicules and, by temporally averaging our data, we show that low-resolution or low-cadence spectropolarimetry can highly underestimate the magnetic field strength. Finally, Sect.~\ref{conclusions} provides the concluding remarks.

 \section{Observations} \label{sec:obs}


The observations used in this work were obtained with the CRisp Imaging SpectroPolarimeter \citep[CRISP;][]{Scharmer_2008} at the Swedish 1-m Solar Telescope \citep[SST;][]{sst} on June 2016 and consist of imaging spectropolarimetry in the \ion{Ca}{II} 8542 \AA\ line with a pixel scale of 0.057\arcsec~pixel$^{-1}$. The spectral resolution was 0.05~\AA\ at line centre and 0.25~\AA\ at the wings, with a total of either 15 or 21 spectral positions, depending on the data set (see Table~\ref{table1}). Data were reconstructed with the Multi-Object Multi-Frame Blind Deconvolution \citep[MOMFBD;][]{VanNoort2005}. The CRISP data reduction pipeline \citep{2015delacruz} was used for additional data reduction, including the cross-correlation method of
\citet{Henriques2012A}.
  
The main targets of the different observations were active regions, or their surroundings, near the solar limb. The data sets thus contain spectropolarimetric observations of both on-disk and off-limb chromospheric structures, both in quiet Sun and in active regions (see Table~\ref{table1} for a summary). Two active regions were under study throughout the three observational days referenced here. On 02 June 2016, NOAA AR 12551 was approaching the west solar limb, finishing its transition on 03 June 2016. It was the subject of study of both observational days, enabling a comprehensive study of the magnetic topology of spicules both near and away from sunspots. On 09 June 2016, AR 12553 was observed as it was emerging from the east solar limb. In all data sets, off-limb spicules are near an active region, while on-disk spicules are in a quiet Sun region. The only exception is the data set from 02 June 2016, in which off-limb and on-disk spicules are respectively in the quiet Sun and near an active region. The disk region of this data set is dominated by a sunspot and, given that our aim is to study the magnetic field of spicules, we ignore the on-disk pixels of 02 June 2016.

Figure~\ref{Figure:1} contextualises the observations, with an image from 03 June 2016. It is compared to a simultaneous image taken with NASA's Solar Dynamics Observatory (SDO) in the Atmospheric Imaging Assembly \citep[AIA;][]{Lemen2012} 171 \AA\ channel. Figure~\ref{Figure:2} shows the four Stokes parameters from one of the 03~June 2016 data sets. The spectropolarimetric data reveal a strong circular polarisation (Stokes~$V$) signal, specially off-limb, where there is less signal superposition. The linear polarisation (Stokes $Q$ and $U$) is weaker and these data are very noisy.

\begin{table*}
  \caption{Summary of the observational data sets.}
  \label{table1}
  \centering
  \begin{tabular}{cccccccc}
    \hline\hline
    Data set & Starting time & AR/QS & Limb / Disk & Cadence [s] & Wavelength & Spectral & Frames \\
    & & & & & range [\AA] &positions& \\
    \hline
    1&2016-Jun-02 07:23 & QS & Limb & 36.33& [-1.5, 1.5] & 15&119 \\
    \hline
    2& & AR &Limb &   &  && \\
    &2016-Jun-03 07:26      & & &36.33 &[-1.5, 1.5]&15 &46 \\
    3&& QS&Disk& &&&\\
    \hline
    4& & AR& Limb &   &  && \\
    &2016-Jun-03 08:17      & & &36.33&[-1.5, 1.5] &15 &36 \\
    5&& QS&Disk& &&&\\
    \hline
    6& & AR&Limb  &   &  && \\
    &2016-Jun-03 08:41     & & &36.33&[-1.5, 1.5] &15 &34 \\
    7&& QS&Disk& &&&\\
    \hline
    8& & AR& Limb  &  &  && \\
    &2016-Jun-03 16:27     & & &36.33 & [-1.5, 1.5]&15 &100 \\
    9&& QS&Disk& &&&\\
    \hline
    10& & AR&  Limb&   &  && \\
    &2016-Jun-09 07:39     & & &26.47 &[-2.0, 2.0]&21 &113 \\
    11&& QS&Disk& &&&\\
    \hline
  \end{tabular}
\end{table*}

To make a quick reference to the data sets, we have labelled each of them with a number (see Table~\ref{table1}). Since we have analysed separately on-disk and off-limb pixels, each data set has been split into two, so that a different number has been assigned in Table~\ref{table1} to off-limb and on-disk data. Adding all data sets, we have 448 spectropolarimetric scans in the \ion{Ca}{II} 8542 \AA\ line, each of them containing some $1000\times 1000$ pixels.

  \section{Data Analysis} \label{sec:Data}
  
  \subsection{The Weak-Field Approximation (WFA)}
In order to infer the line-of-sight (LOS) magnetic field component, $B_{\mathrm{LOS}}$, the Weak Field Approximation \citep[WFA;][]{landi,Centeno2018,kuridze2019} is used. The WFA is a simplification of the radiative transfer equations as a function of the Stokes parameters that can be applied when some conditions are met.
  
  The first condition is that the Zeeman width ($\Delta\lambdaB$) needs to be much smaller than the Doppler width ($\Delta\lambdaD$):
\begin{center}
  \begin{equation}
      \overline{g} \frac{\Delta\lambdaB}{\Delta\lambdaD} \ll 1,
  \label{eq:1}
  \end{equation}
\end{center}
where $\overline{g}$ is the effective Land\'e factor of the line \citep{Landi1982}, which for the \ion{Ca}{II} 8542 \AA\ line is 1.1. 
The above inequality can be written as

\begin{center}
  \begin{equation}
      \frac{1.4 \times 10^{-7}\lambda_{0} \overline{g} B}{\sqrt{1.663\times 10^{-2} \frac{T}{\mu} + \xi^2}} \ll 1,
  \label{eq:2}
  \end{equation}
\end{center}
where $T$ is the temperature (in K), $\lambda_0$ is the wavelength of the transition (in \AA), $\mu$ is the atomic weight of the species, $\xi$ is the microturbulent velocity (in km $\mathrm{s}^{-1}$) and $B$ is the magnetic field intensity (in G). The second condition is that the magnetic field must be uniform in the regions of line formation along the line-of-sight. Following the derivation presented by \citet{landi}, a perturbative scheme can be applied to the radiative transfer equations when such conditions are met and a relation between the Stokes $I$ and $V$ parameters can be derived:

\begin{center}
  \begin{equation}
      V (\lambda) = -4.67 \times 10^{-13} \overline{g} f \lambda_{0}^{2} B_{\mathrm{LOS}} \frac{\partial I (\lambda)}{\partial \lambda},
  \label{eq:3}
  \end{equation}
\end{center}
where $f$ is the magnetic filling factor \citep{landi,AsensioRamos2011}, which here is assumed to be unity at chromospheric heights due to the expansion of the magnetic field. As pointed out by \cite{Centeno2018}, however, $f=1$ might not be true everywhere in the Sun \citep{2007A&A...469..721S} and one must keep this in mind when analysing the results presented here. After properly assessing the applicability of the theory, the LOS magnetic field component can be inferred from the observed Stokes profiles with Eq.~(\ref{eq:3}).

\subsection{Weak Field Approximation conditions}
In order to properly make use of the WFA and to ensure its validity, two restrictions on the Stokes $I$ and $V$ profiles are imposed on every pixel of each data set. If one of the restrictions is not met, the pixel is not used in the analysis and its $B_{\mathrm{LOS}}$ is not calculated.

The first condition directly comes from Eq.~(\ref{eq:2}). Substituting the numerical values for the \ion{Ca}{II} 8542 \AA\ line, a temperature of 7500 K and a microturbulent velocity\footnote{This is the most restrictive microturbulent velocity value we have found in the literature for the \ion{Ca}{II} 8542 \AA\ line. Figure~11 of \citet{Vernazza1981} gives $\xi=5$~km~s$^{-1}$ at 1500~km height. Both \citet{2013A&A...556A.115D} and \citet{Quintero2016} use $\xi=3$~km~s$^{-1}$. Finally, \citet{Jurcak2018} quote values in the range 4--5~km~s$^{-1}$ and 3.5--13~km~s$^{-1}$ for VAL \citep{Vernazza1981} and FAL \citep{Fontenla1993} models, respectively.} of 3 km s$^{-1}$ results in the condition
\begin{center}
  \begin{equation}
      B \ll 2650 \hspace{2pt} \mathrm{G}.
  \label{eq:4}
  \end{equation}
\end{center}
\noindent
In principle, magnetic fields of such intensity are not expected to be present at chromospheric heights outside sunspots.

\begin{figure}
   \centering
   \includegraphics[width=8cm]{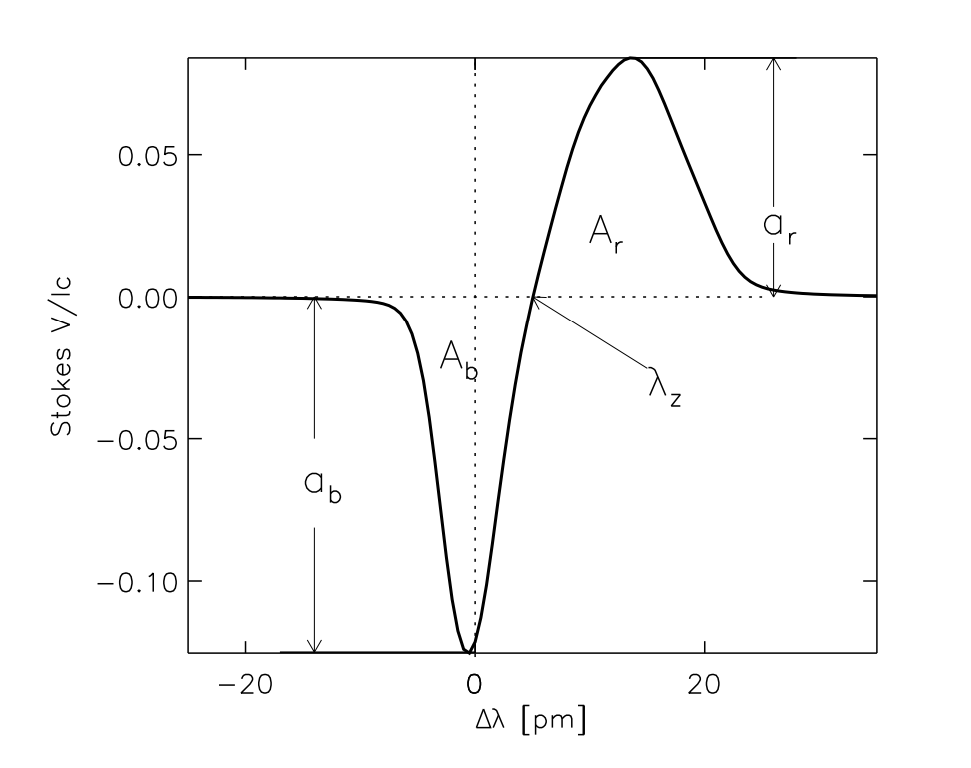}
      \caption{Typical asymmetric Stokes $V$ profile \citep[from][]{sheminova2005}. The meaning of the various quantities is given in the text.}
         \label{Figure:3}
   \end{figure}

We now analyse the requirement that the magnetic field must be independent of optical depth. The presence of a magnetic field gradient along the line-of-sight can leave footprints in the profiles of the circularly polarised radiation \citep{2001sigwarth,sheminova2005,1993solanki}. In a static atmosphere in LTE, the usual profile of the Stokes $V$ parameter as a function of wavelength is characterised by two lobes of equal area and amplitude, with a zero crossing at the line center, $\lambda_0$ \citep{auer}. A typical asymmetric Stokes $V$ profile is shown in Fig.~\ref{Figure:3}, in which several parameters used to quantify the Stokes $V$ asymmetry are defined. Amplitude and area asymmetries of the circularly polarised radiation, $V$, are strongly linked to variations in the magnetic field configuration combined with velocity shifts along the line-of-sight. Furthermore, velocity gradients along the LOS also give rise to asymmetries in Stokes~$I$ \citep{Kuridze2015}. This means that they also produce area and amplitude asymmetries in the Stokes~$V$ lobes. Then, very asymmetric circular polarisation signals must be filtered out. The presence of noise in the data can also give rise to Stokes $V$ asymmetries, which further complicates the analysis. Hence, a careful assessment of the asymmetry of the observed profiles must be devised before applying Eq.~(\ref{eq:3}).

The definition of area and amplitude asymmetries used here are, respectively,
\begin{center}
  \begin{equation}
\delta A =\frac{|A_b -A_r|}{\max(|A_b|,|A_r|)}, 
  \label{eq:5}
  \end{equation}
\end{center}
and
\begin{equation}
\delta a = \frac{|a_b - a_r|}{\max(|a_b|,|a_r|)},
\label{eq:5b}
\end{equation}

\noindent where $A_b$ and $A_r$ are the respective areas of the Stokes~$V$ blue and red lobes, whereas $a_b$ and $a_r$ are the respective amplitudes of the Stokes~$V$ blue and red lobes; see Fig.~\ref{Figure:3}.

\begin{figure*}
   \centering
   \includegraphics[width=8cm]{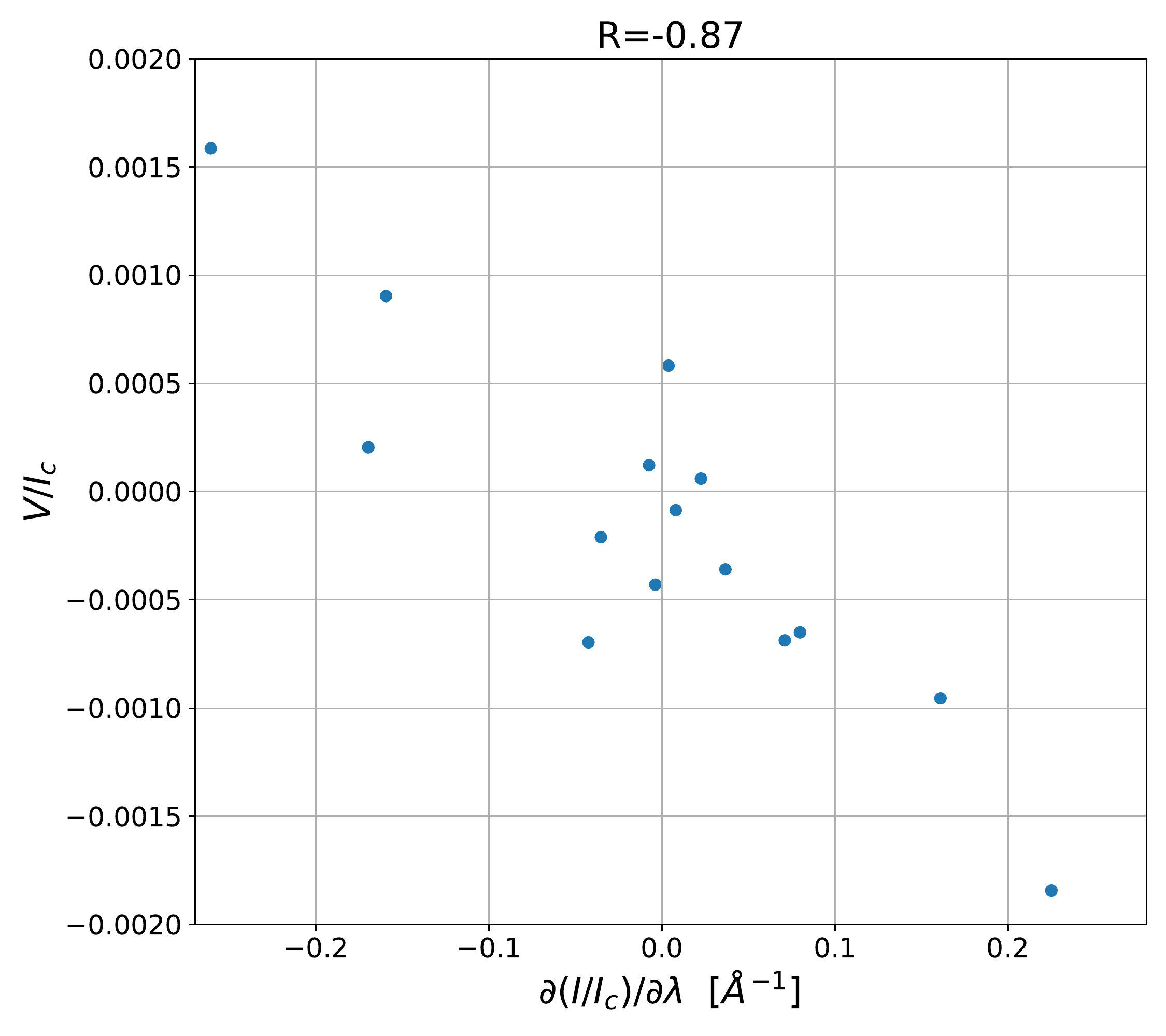}
   \includegraphics[width=8cm]{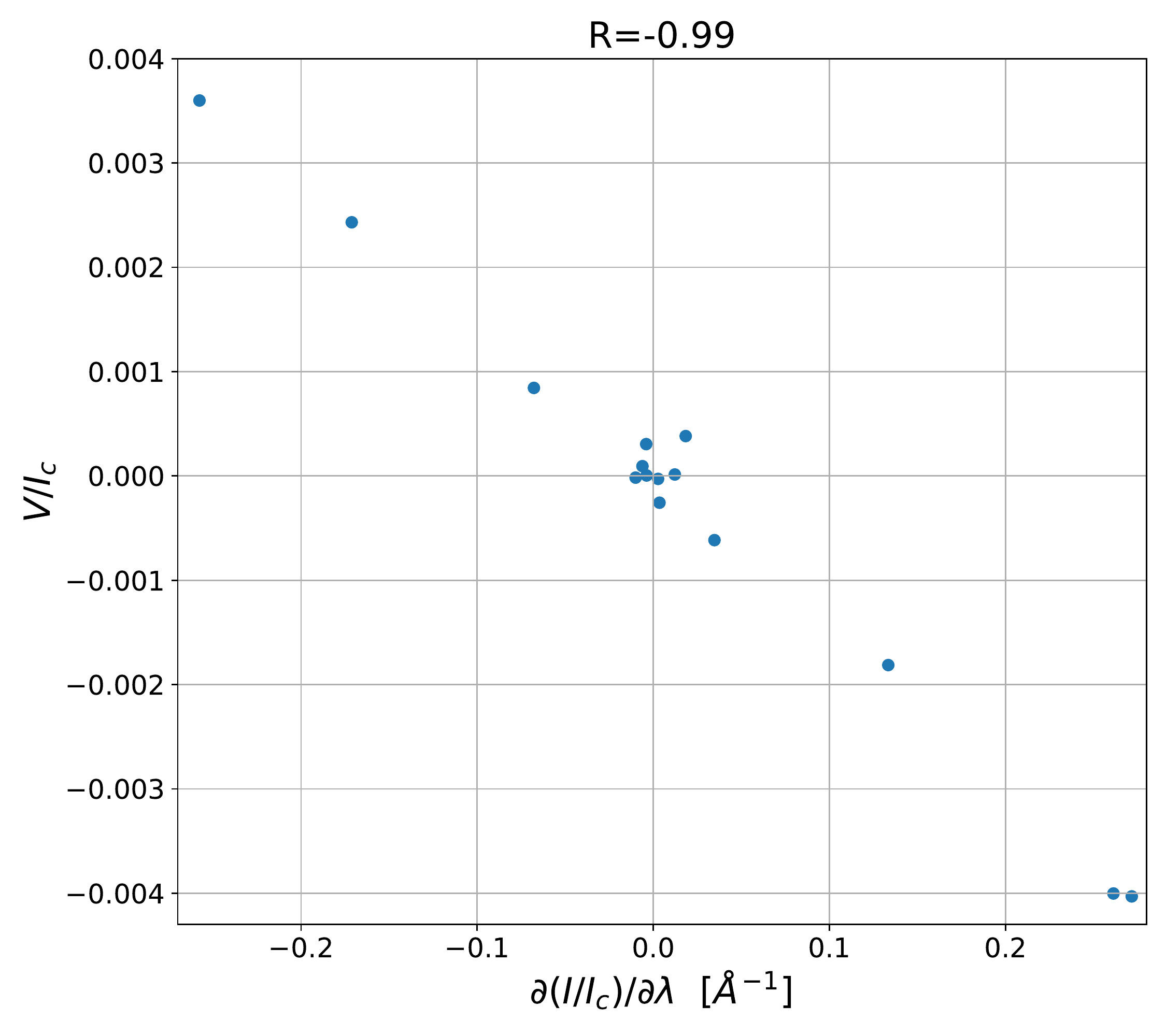}
   \caption{Plot of Stokes $V$ against $\partial I/\partial \lambda$ for a pixel in which |$R$| $<$ 0.9 (left) and a pixel in which |$R$| $>$ 0.9. The corresponding values of $R$ are shown on top of each panel.}
              \label{Figure:4}%
\end{figure*}
    
\begin{figure*}
   \centering
   \includegraphics[width=8cm]{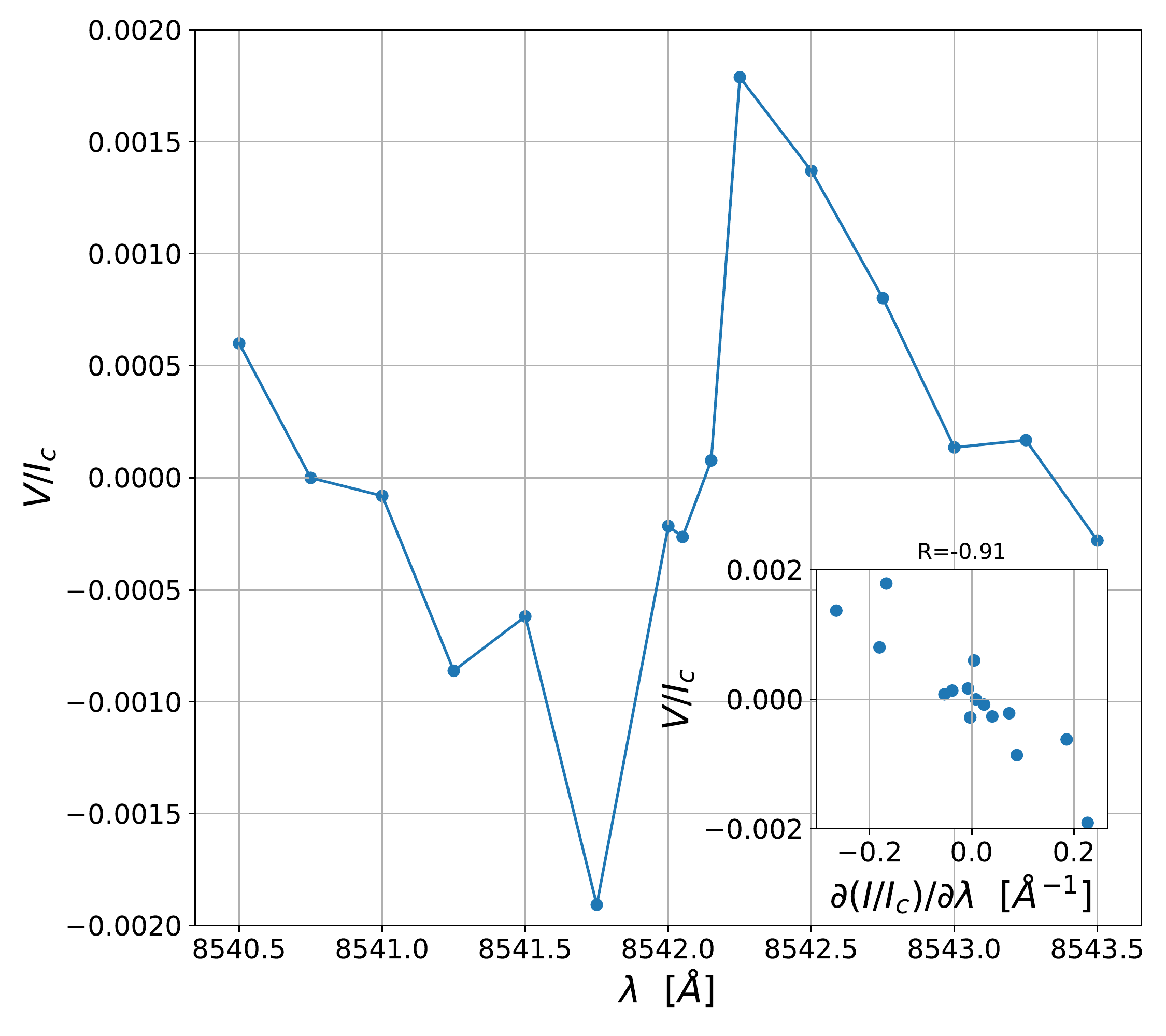}
   \includegraphics[width=8cm]{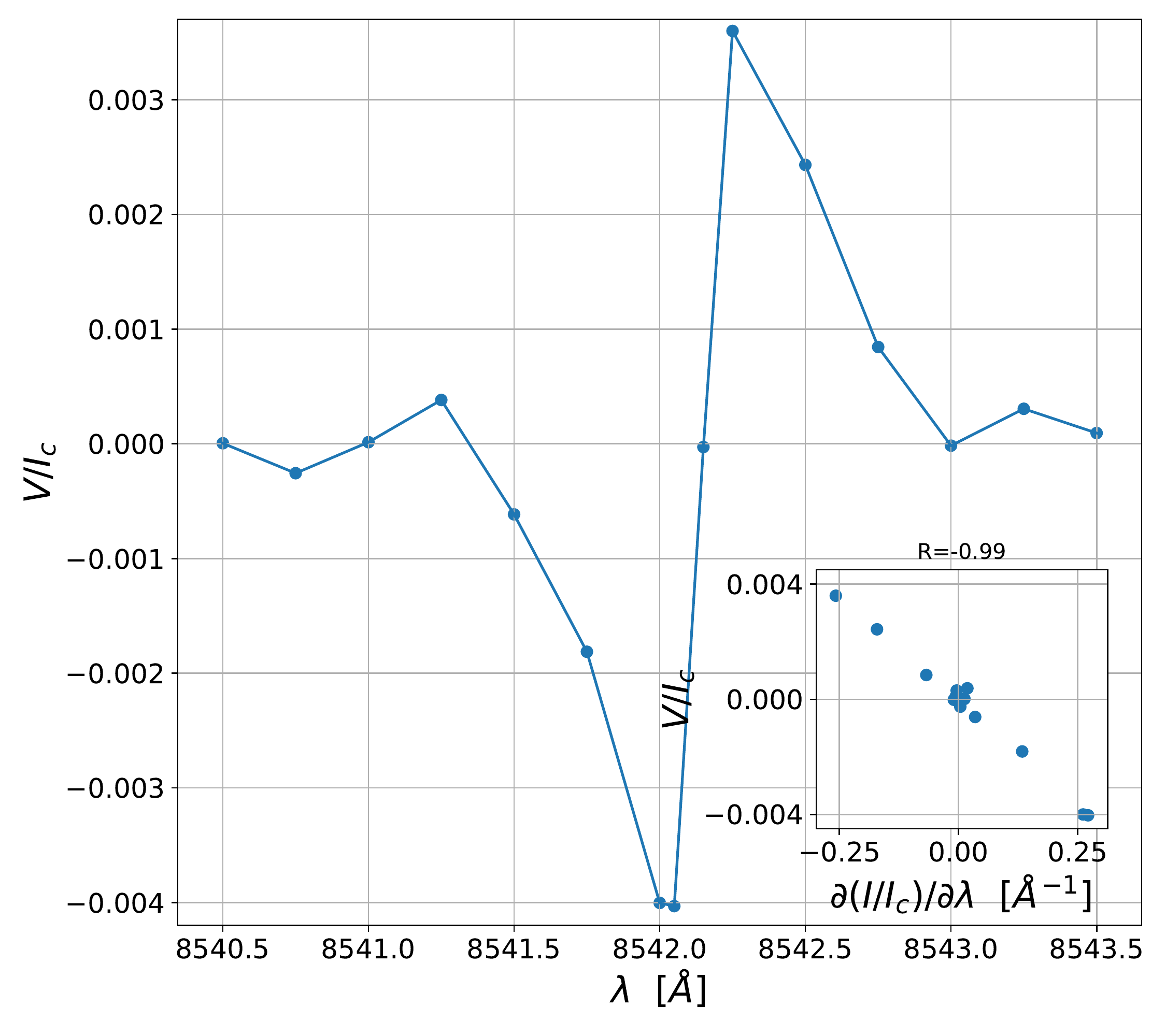}
   \caption{Examples of Stokes $V$ profiles in which the correlation criterion $|R|>0.9$ is met. Left: the asymmetry criterion is not satisfied. Right: the asymmetry criterion is satisfied. The insets show the plots of Stokes $V$ vs $\partial I/\partial \lambda$, both of which have passed the filter $|R|>0.9$.}
              \label{Figure:5}%
\end{figure*}

\begin{figure*}
   \centering
   \includegraphics[width=8cm]{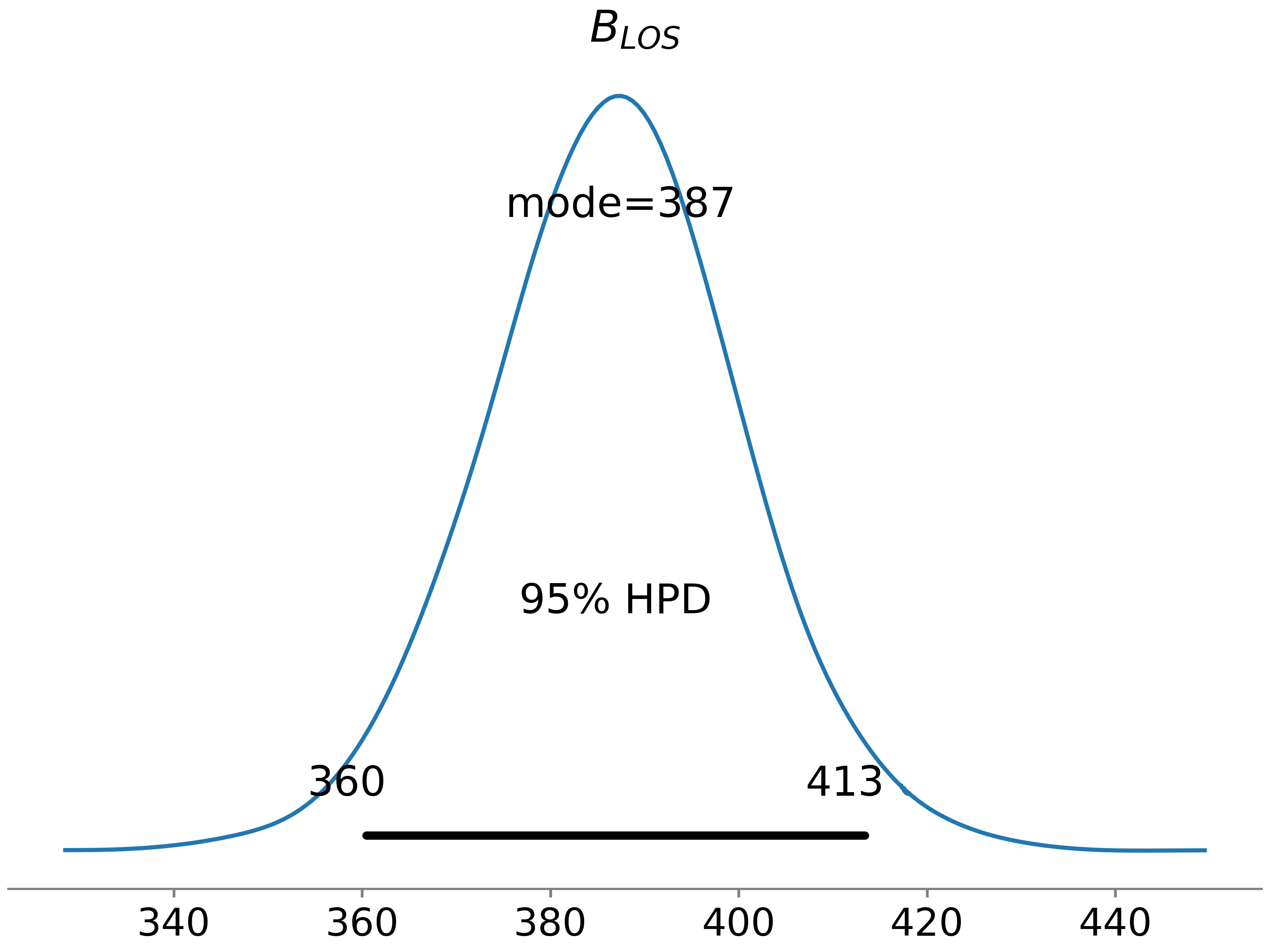}
   \includegraphics[width=8cm]{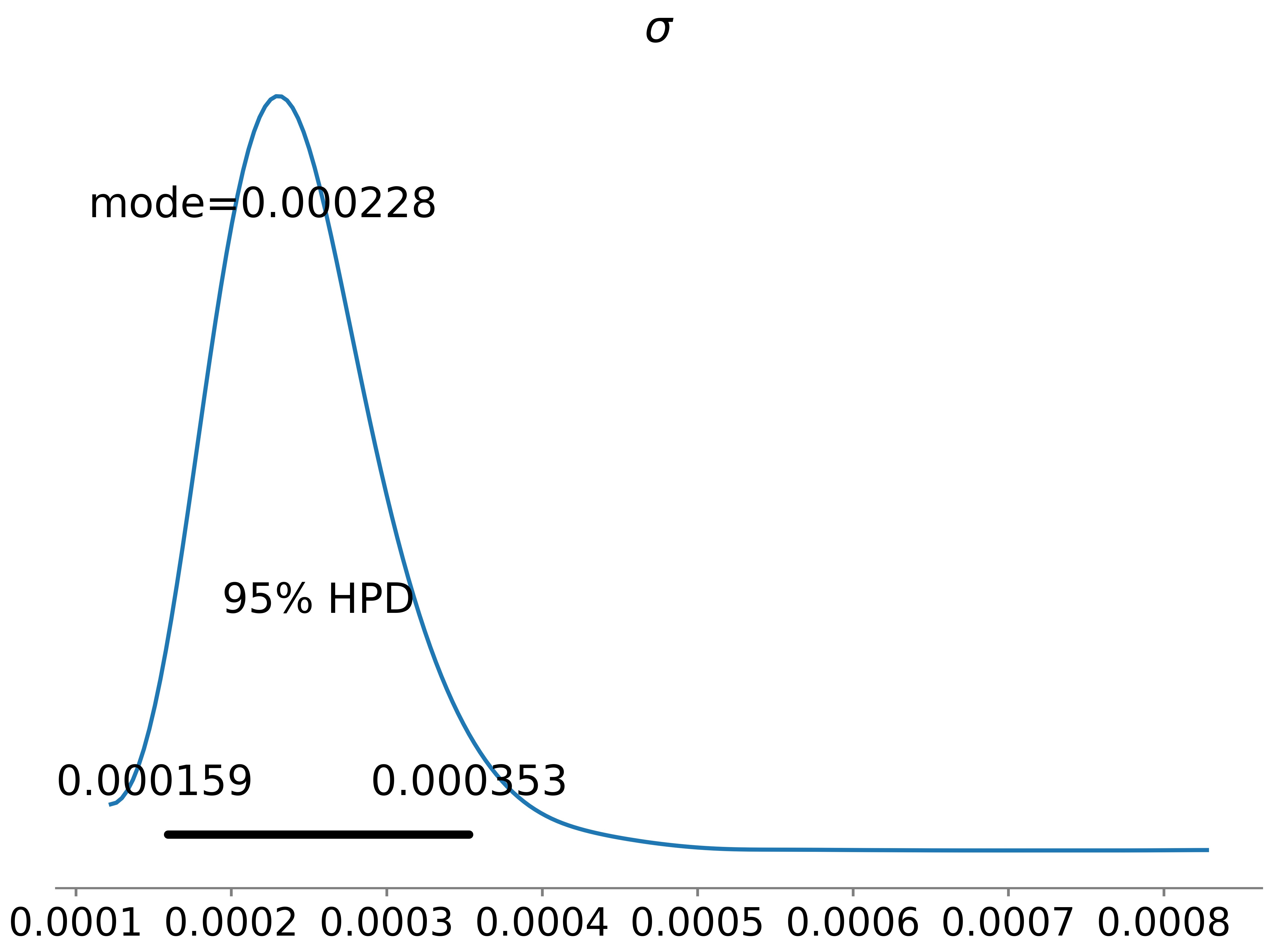}
   \caption{Results of the Bayesian inference for one pixel. The mode of the posterior distribution and the 95\% HPD interval of $B_{\mathrm{LOS}}$ (left; in G) and $\sigma$ (right; in normalised count units) are shown.}
              \label{Figure:6}%
\end{figure*}

In order to control the asymmetry of the Stokes $V$ profile, the following condition is imposed:
\begin{equation}
 \delta a < \frac{\varepsilon \eta}{\max(|a_b|,|a_r|)}.
\label{eq:6}
\end{equation}
$\varepsilon$ is a number between 0 and 1 (we take\footnote{To see how much asymmetry is allowed by this value, see Fig.~\ref{Figure:5}.} $\varepsilon=0.5$)  and $\eta$ is the average amplitude of the noise present in the specific time instant of the data set under consideration. It is estimated by averaging the signal far off-limb where only noise is expected to be present.
In our analysis we have seen that almost every time that condition~(\ref{eq:6}) is met, the area asymmetry is smaller than the amplitude asymmetry. Hence, no restriction is imposed on $\delta A$.

\subsection{Method}

In this work we use a Bayesian approach to infer parameters at each pixel and each time, but before applying this technique we must ensure that the Stokes $I$ and $V$ profiles are suitable for the WFA. The first step is based on the fact that Eq.~(\ref{eq:3}) establishes a linear relation between $V(\lambda)$ and $\partial I/\partial \lambda$. The Pearson correlation coefficient \citep{pearson}, $R$, between such quantities is calculated in every pixel, and only those for which |$R$| > 0.9 are selected. Figure~\ref{Figure:4} shows a comparison between a profile that meets such criterion and a profile which does not, evidencing the need of imposing this constraint on $R$. 

After pixels for which the correlation criterion does not hold have been filtered out, the next step is to analyse the asymmetry of the Stokes $V$ profile. Condition~(\ref{eq:6}) requires a precise measurement of $a_b$ and $a_r$. This is troublesome when dealing with our data sets, since only 15 or 21 spectral positions are sampled, so that the actual extrema of $V$ will most likely not fall exactly at one of the sampled wavelengths. For this reason, and also because our measurements are affected by noise, two models are fitted to the observed profiles in order to determine their asymmetry more accurately. The first model is a simple sum of Gaussian functions, with 6 free parameters, while the second model is the derivative of a skewed Gaussian function, with 4 free parameters to adjust. Both models allow for the existence of asymmetries. The Bayesian information criterion \citep[BIC;][]{sch} is used to identify the model that best fits the Stokes $V$ profile at each pixel. The value of $\delta a$ is established from the fitted profile and condition~(\ref{eq:6}) is checked. An example of $V(\lambda)$ profile that does not meet the asymmetry criterion is shown in the left panel of Fig.~\ref{Figure:5} alongside one that does meet such criterion in the right panel.

\begin{figure*}
   \centering
   \includegraphics[width=18cm]{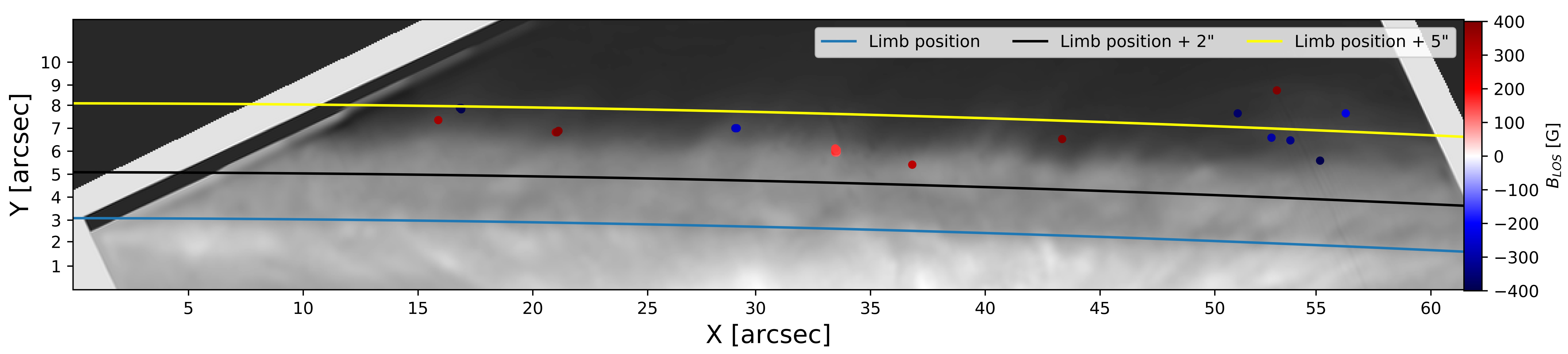}
   \includegraphics[width=18cm]{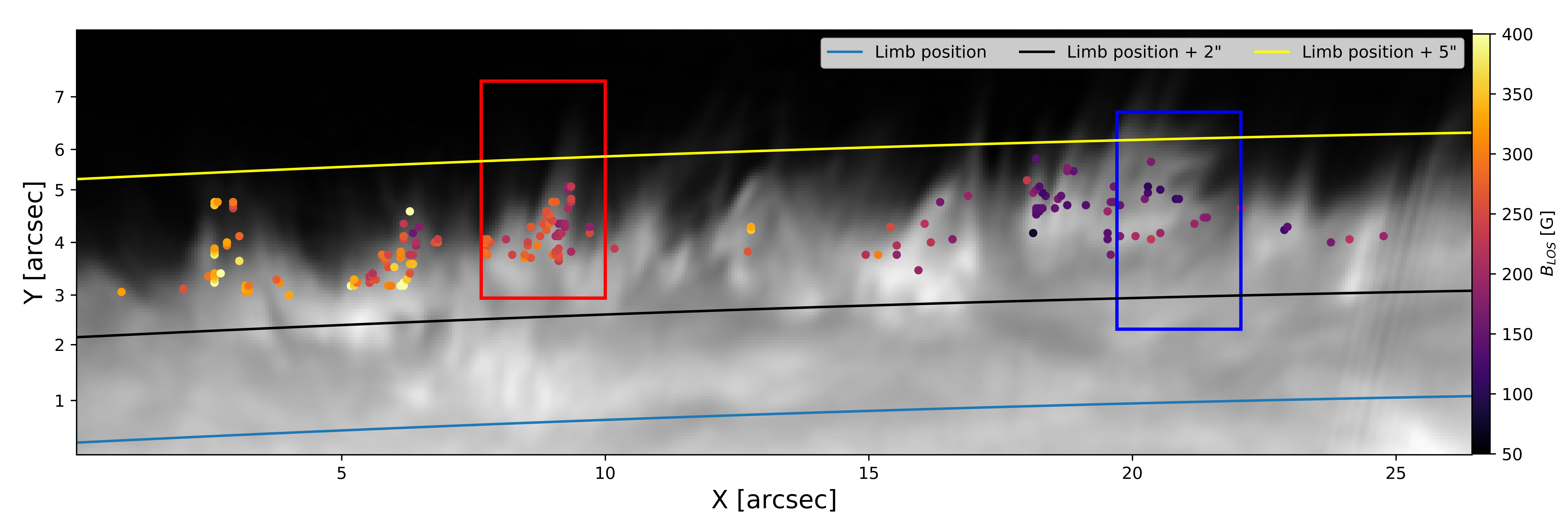}
   \caption{Results from the Bayesian inference for the off-limb data of 02 June 2016 at 07:46:06 UT, belonging to data set \#1 (top frame), and 03 June 2016 at 16:43:20 UT, belonging to data set \#8 (bottom frame). The points are coloured based on the mode from the posterior distribution of the LOS magnetic field intensity (see colour bar at the right) and are plotted on top of the \ion{Ca}{II} 8542 \AA\ line centre intensity image. The blue line represents the position of the solar limb, while the black and yellow curves represent the heights 2\arcsec\ and 5\arcsec\ above the limb, respectively. The red and blue rectangles in the bottom panel are used in Fig.~\ref{Figure:9}.}
              \label{Figure:7}%
    \end{figure*}       

%
%
%


\begin{figure*}
   \centering
   \includegraphics[width=6cm]{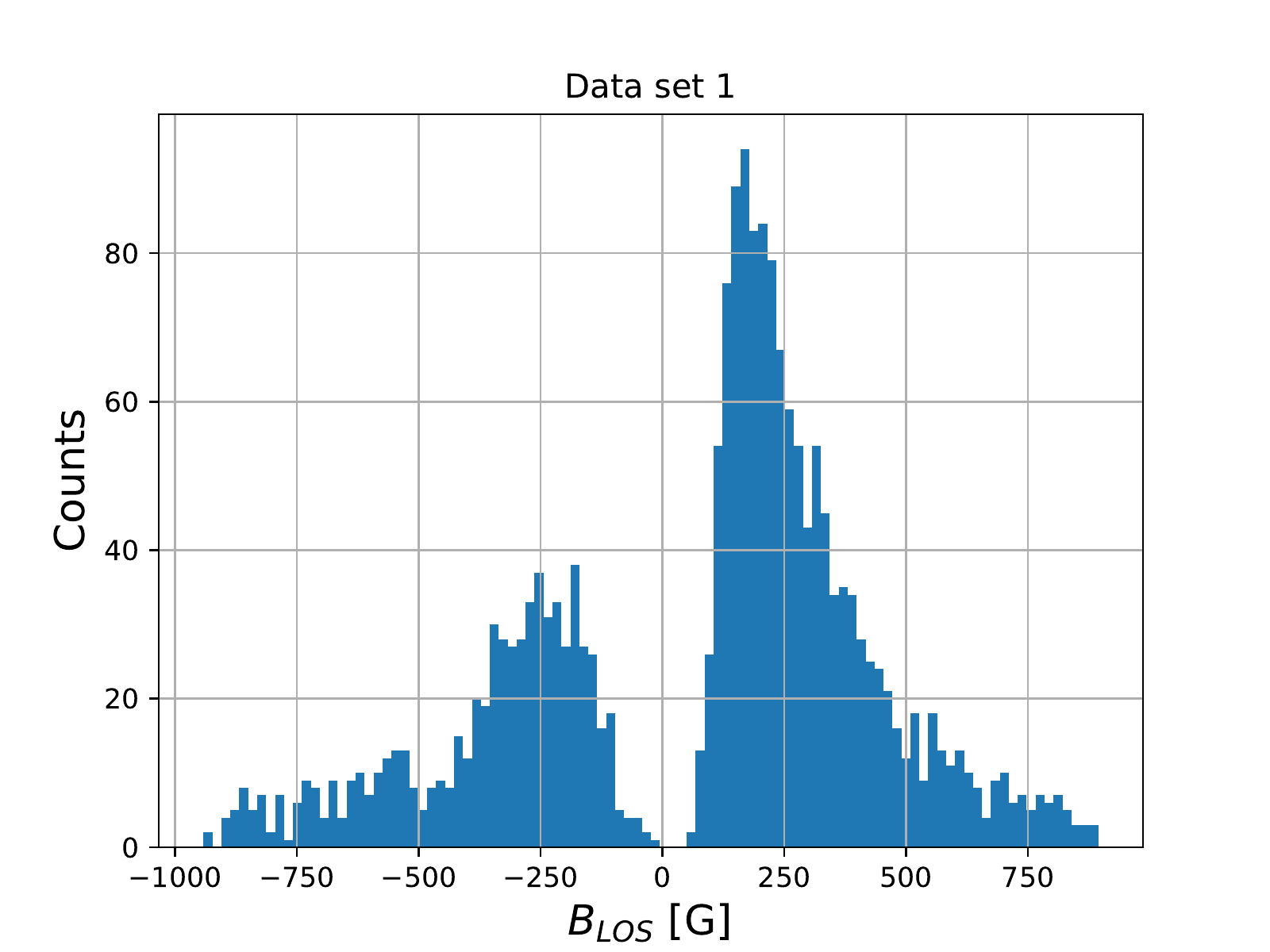}
   \includegraphics[width=6cm]{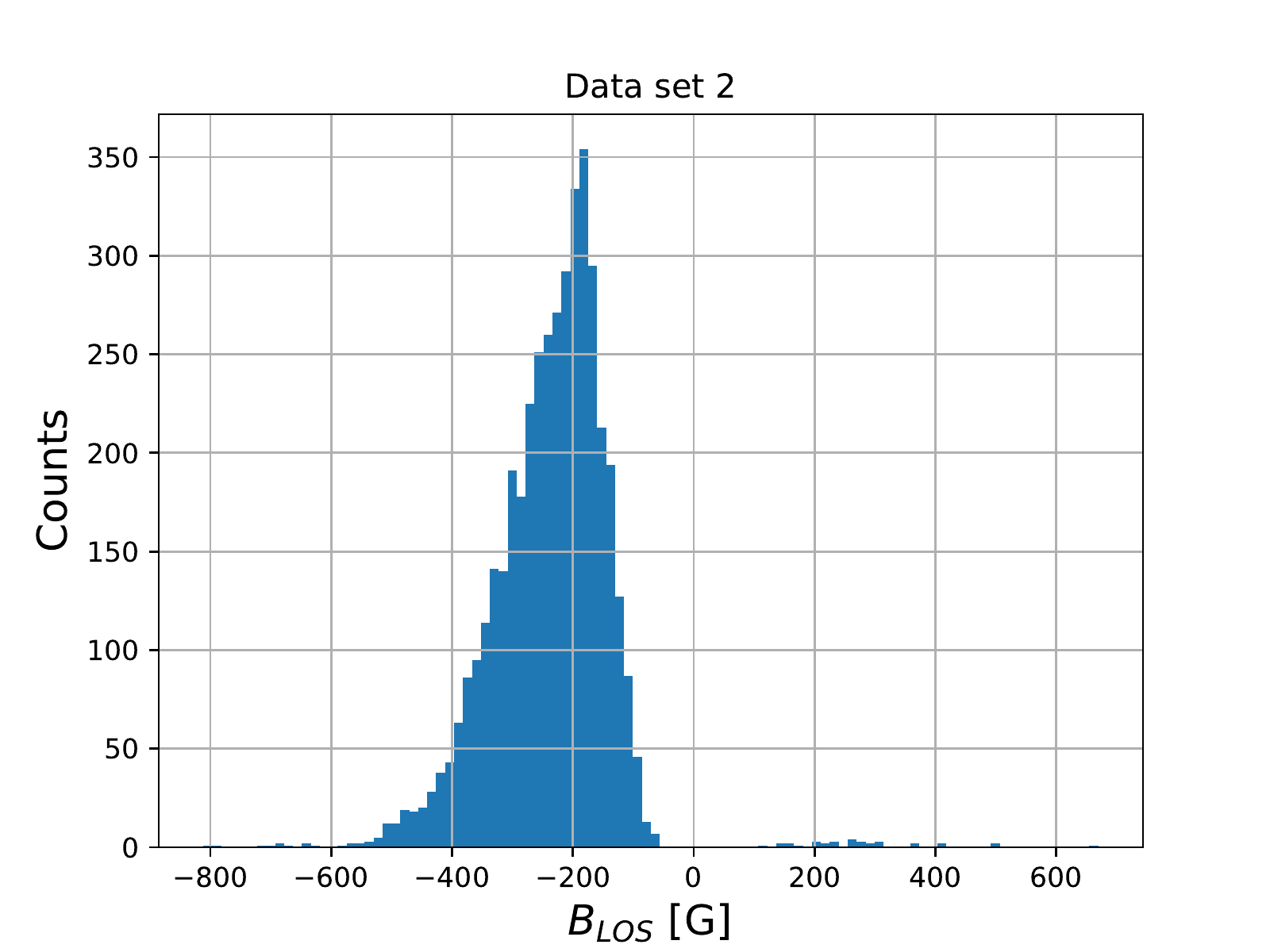}
      \includegraphics[width=6cm]{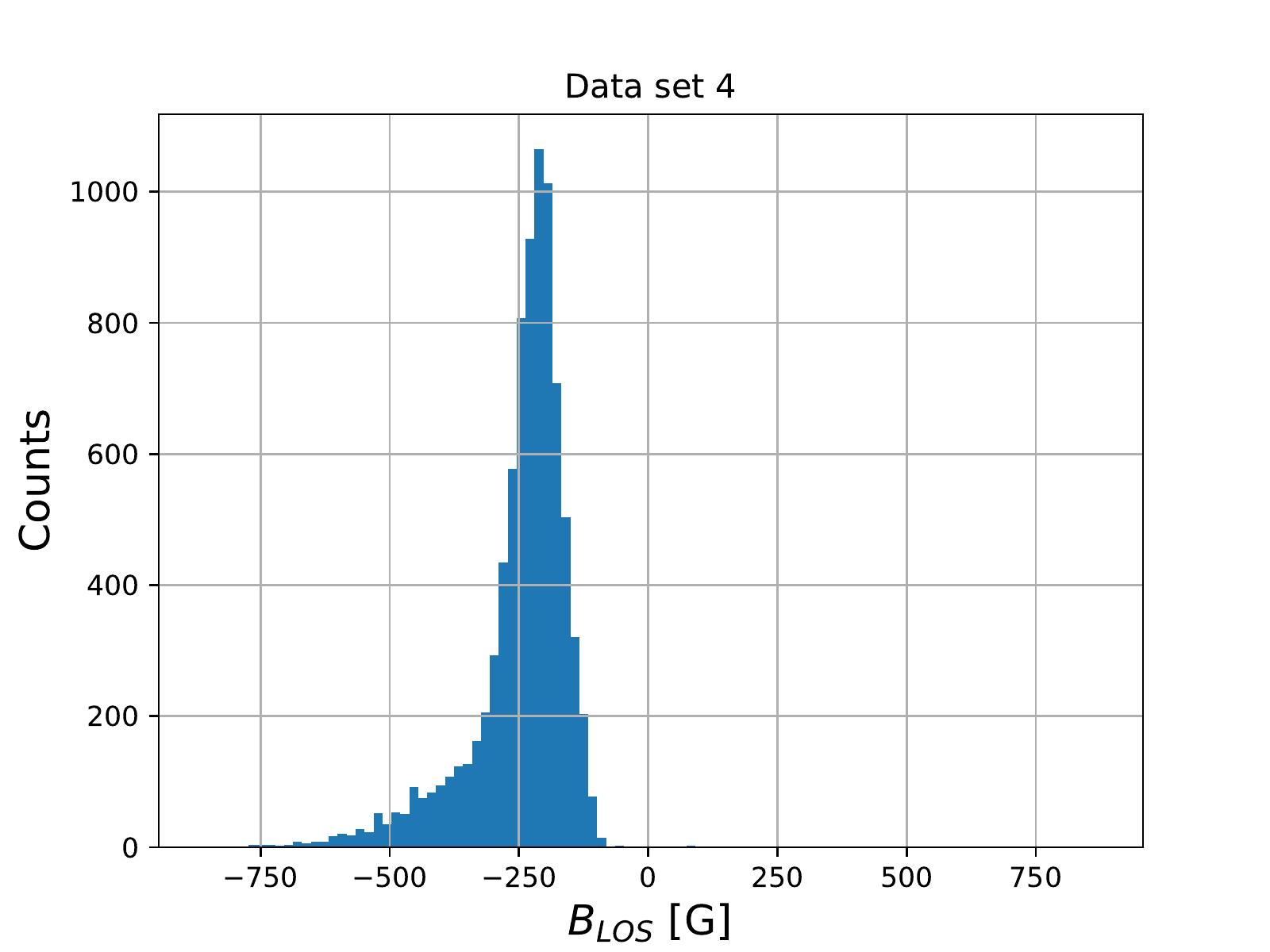}
   \includegraphics[width=6cm]{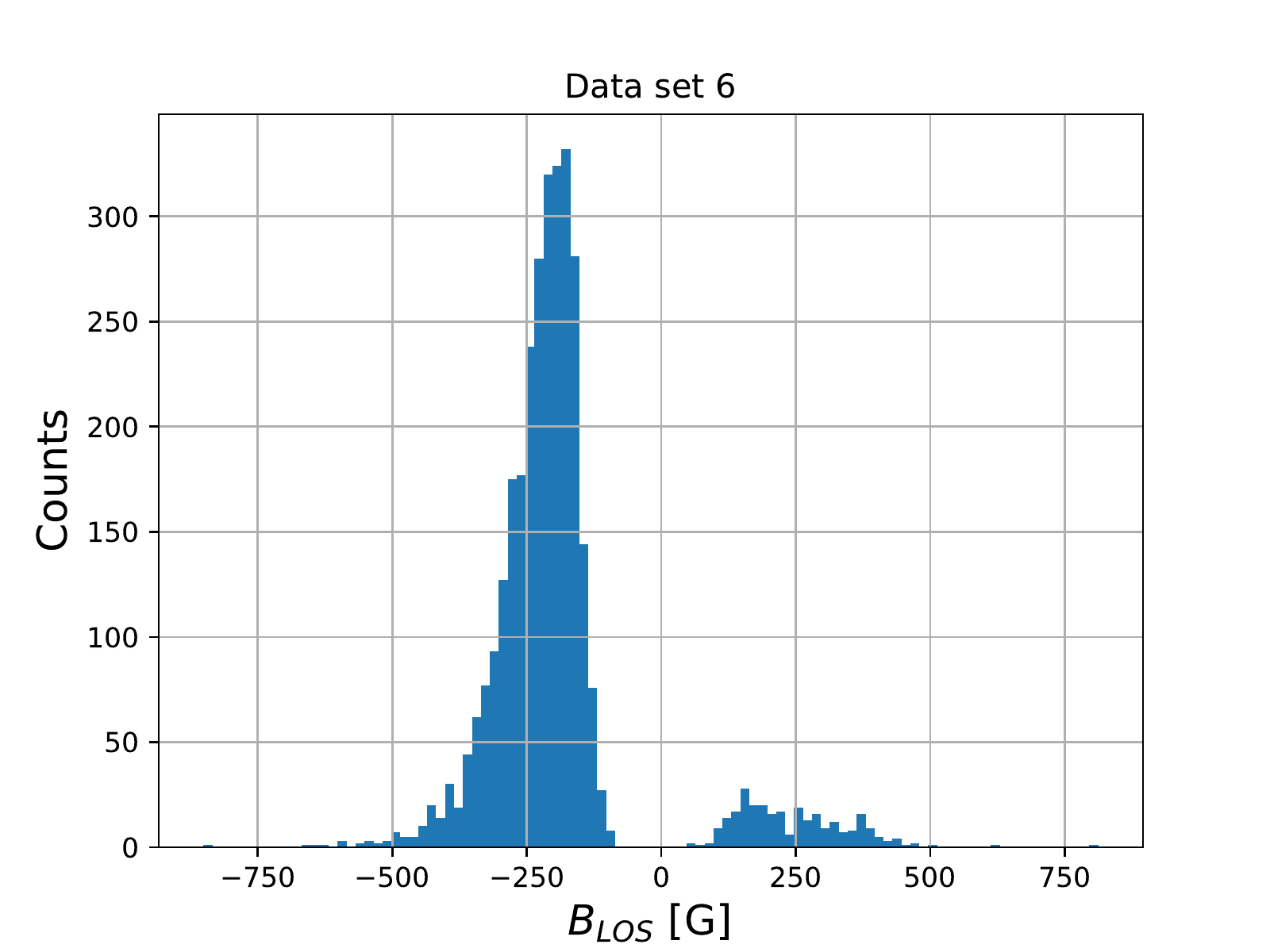}
      \includegraphics[width=6cm]{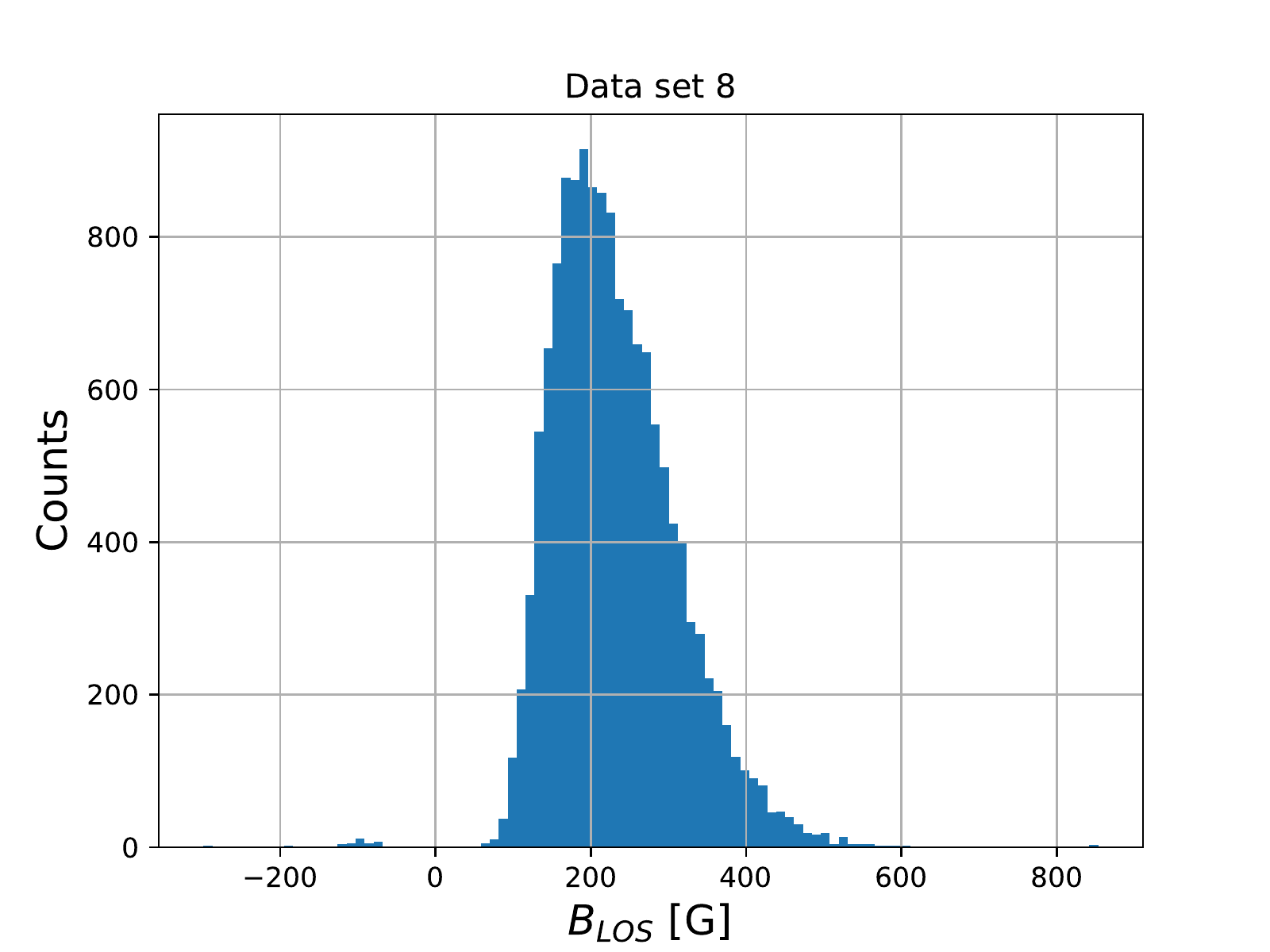}
   \includegraphics[width=6cm]{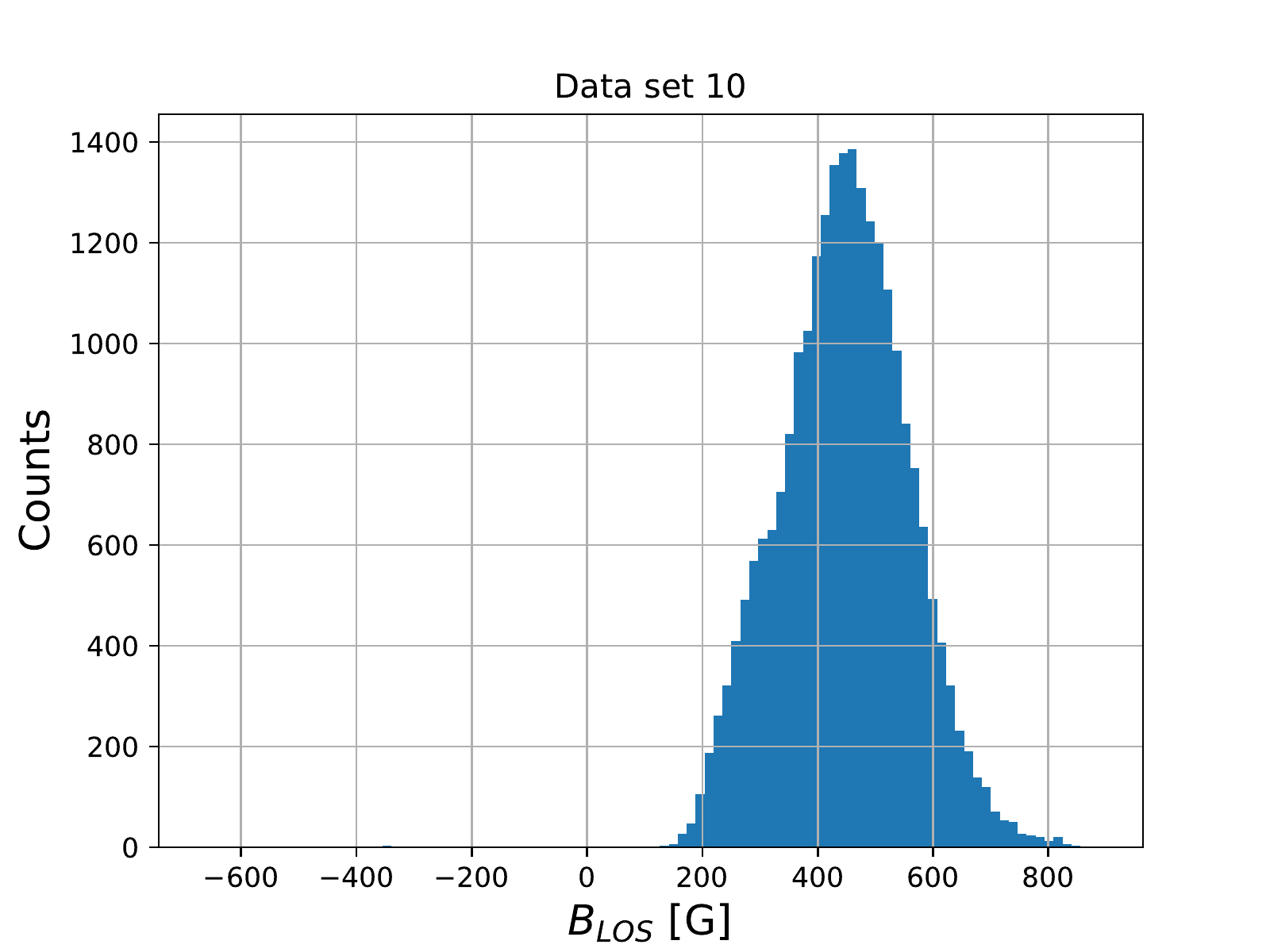}
   \caption{Histogram of  $B_{\mathrm{LOS}}$ values for the off-limb pixels. The data set labels are those given in Table \ref{table1}.}
              \label{Figure:11}%
    \end{figure*}    

\begin{figure*}
   \centering
   \includegraphics[width=15cm, trim=1px 0 0 0, clip]{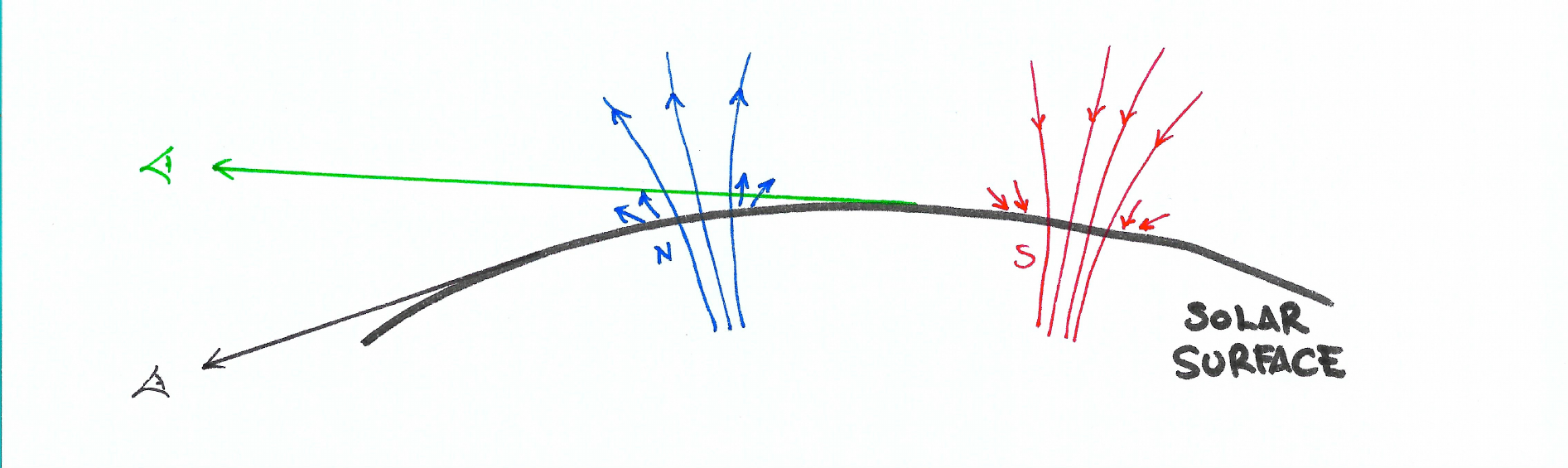}
   \caption{Schematic representation of the magnetic polarity of AR 12551 and the spicule fields (short blue and red arrows) around the dominant magnetic polarities (long blue and red arrows). The green and black straight lines respectively represent the line-of-sight tangent to the solar limb of data sets \#2, \#4, \#6 (morning of 03 June 2016) and of data set \#8 (afternoon of 03 June 2016).}
              \label{Figure:10}%
    \end{figure*}

    \begin{figure*}
   \centering
   \includegraphics[width=6cm]{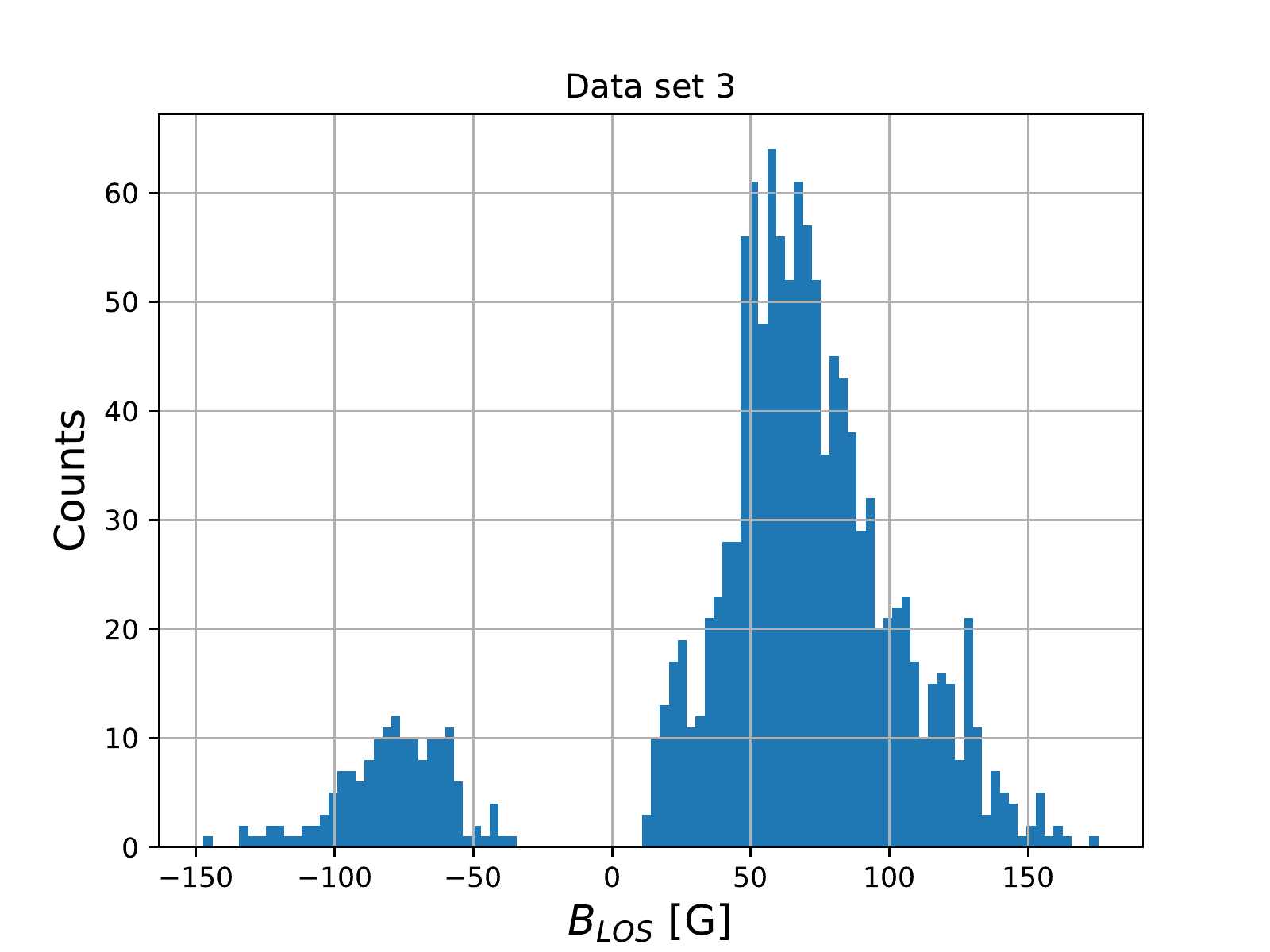}
   \includegraphics[width=6cm]{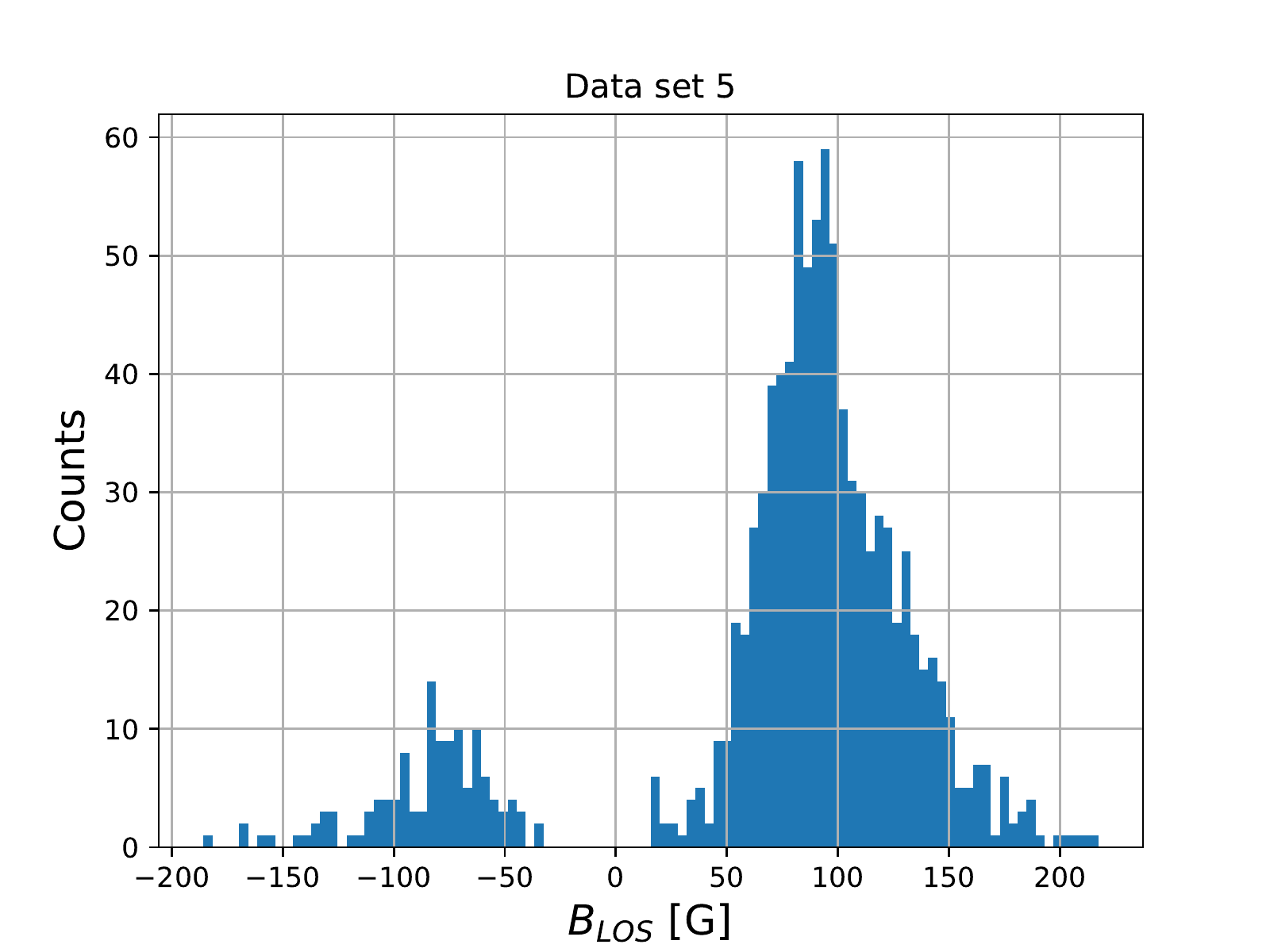}
      \includegraphics[width=6cm]{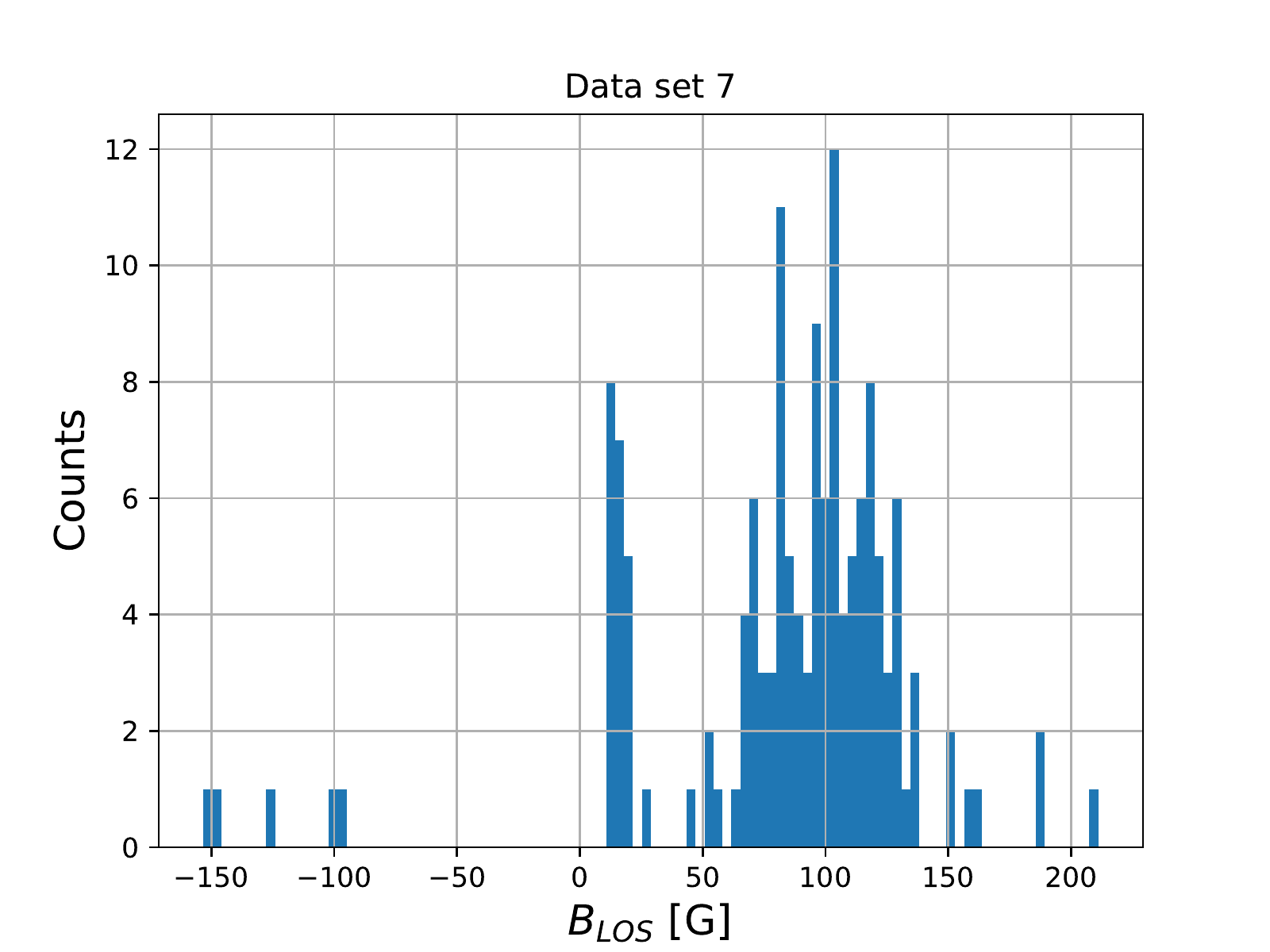}
   \includegraphics[width=6cm]{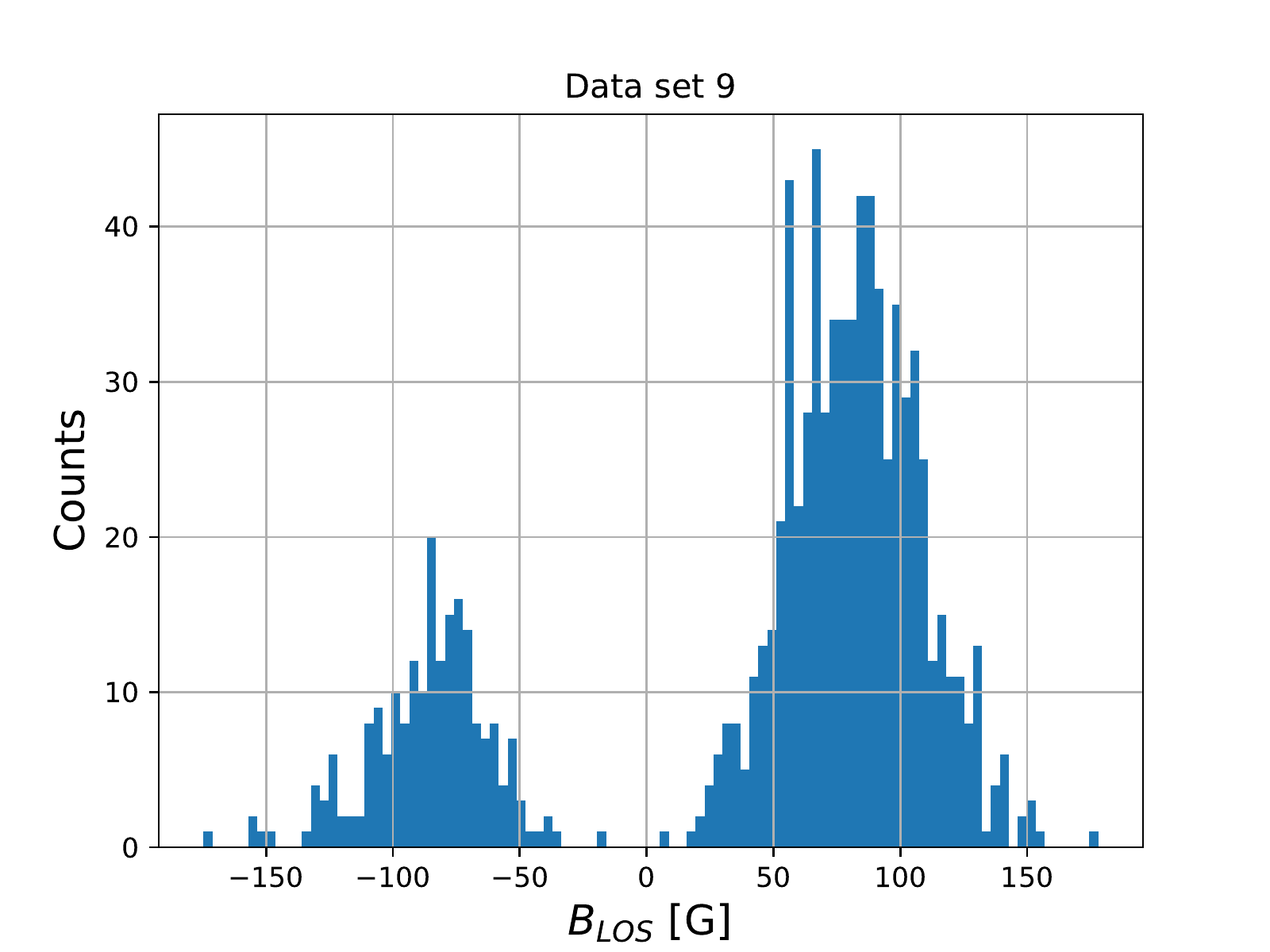}
      \includegraphics[width=6cm]{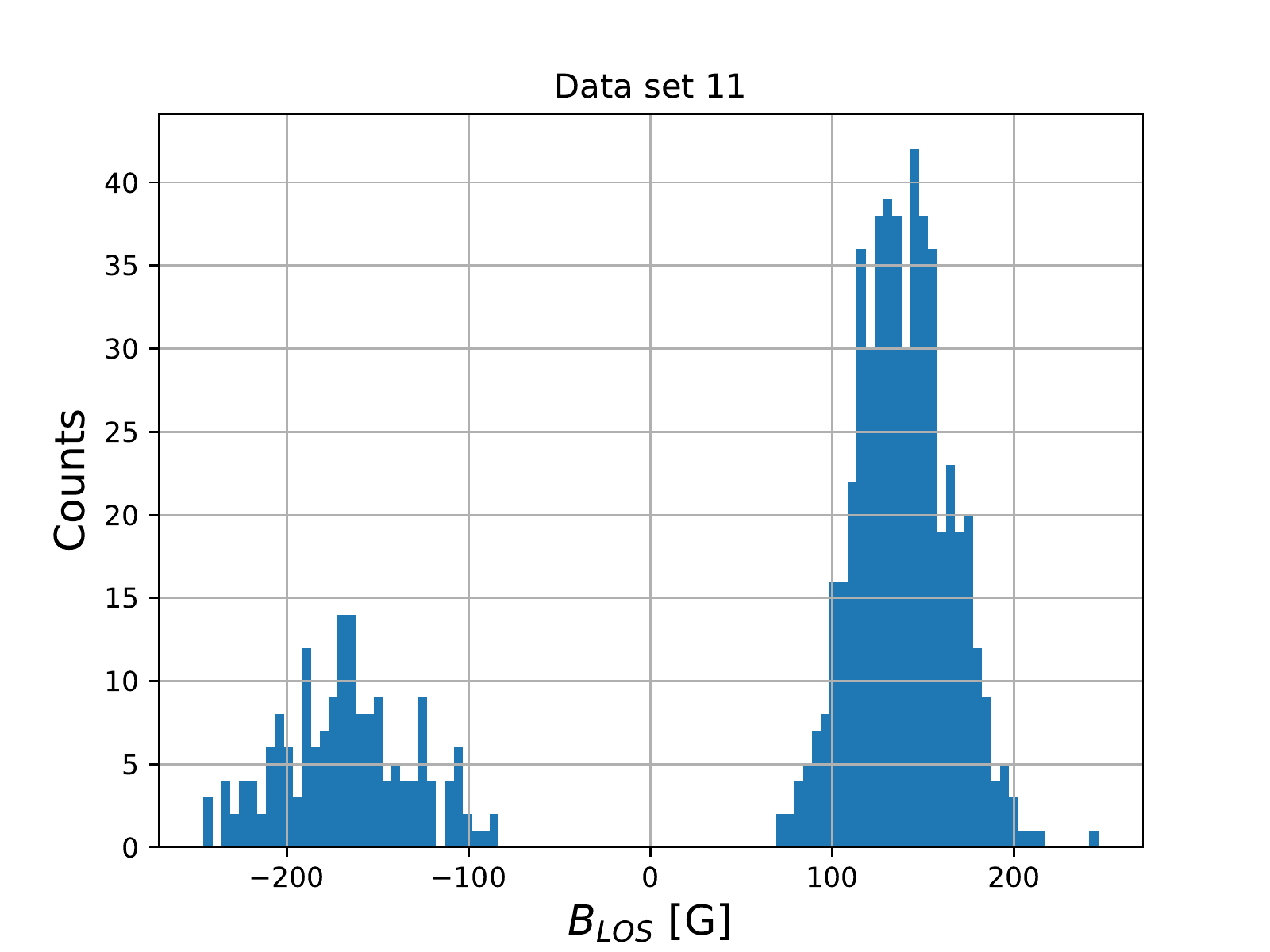}
    \caption{Same as Fig.~\ref{Figure:11} for on-disk pixels. The data set labels are those given in Table \ref{table1}.}
              \label{Figure:12}%
    \end{figure*}

After imposing the correlation and asymmetry criteria to all pixels, we are left with those whose Stokes $I$ and $V$ profiles are suitable for the application of the WFA. We use Bayes' formula

\begin{equation}
P(\mathrm{params}|\mathrm{data})\hspace{0.5cm} \alpha \hspace{0.5cm} P(\mathrm{data}|\mathrm{params})\times P(\mathrm{params}),
\end{equation}

\noindent where $P(\mathrm{params}|\mathrm{data})$ is the posterior distribution, which once obtained gives the probability distribution of the parameters given the observed data (i.e. $I$ and $V$); $P(\mathrm{params})$ is the prior distribution, which reflects the belief one has about the values that the unknown parameters can take before the measurements are performed; and $P(\mathrm{data}|\mathrm{params})$ is the likelihood distribution, which gives the probability distribution of possible values of $V(\lambda)$ and $\partial I/\partial \lambda$ that can be measured for given values of the parameters. In the absence of noise and for a fixed \BLOS, $V(\lambda)$ would be given by the right-hand side of Eq.~(\ref{eq:3}). The presence of noise in the data, however, affects this measurement and for this reason the likelihood distribution is chosen as a Gaussian distribution for $V(\lambda)$ centred at the value given by the right-hand side of Eq.~(\ref{eq:3}) and with a standard deviation $\sigma$. This means that, besides \BLOS, there is a second unknown parameter, $\sigma$. Regarding the prior distribution, the choice made here is to assign a uniform prior to $B_{\mathrm{LOS}}$ centred at the value obtained from the slope, $s$, of the least-squares fit of $V(\lambda)$ vs $\partial I/\partial \lambda$:

\begin{equation}\label{eq:BLOS_slope}
    B_{\mathrm{LOS}}= \frac{s}{-4.67 \times 10^{-13} \overline{g}  \lambda_{0}^{2}},
\end{equation}

\noindent and with a width of 500 G. We have tried other values of the width and have found that this one is suitable: it does not constrain the Bayesian inference too much so as to force it to yield a \BLOS\ in a narrow range, such as would happen for a much smaller width, and does not carry with it the penalty of a higher computation time, such as would be the case for a larger width. A half-Cauchy distribution is chosen as a prior for $\sigma$, fixing the scale parameter $\beta$ to 0.5, so that the distribution has mode 0 and $\sigma$ has 70\% probability of being less than one. The values $\beta=1$ and $\beta=2$ have been used in preliminary tests, yielding identical results but with longer computational time.

In order to carry out the product between the likelihood and the prior distributions, and to obtain the posterior distribution and the marginalisation, a numerical sampler is needed. This is done with PyMC3 \citep{pymc3}, a Python package devoted mainly to Bayesian inference problems. PyMC3 makes use of Markov Chain Monte Carlo methods to sample the posterior, and it also includes a wide functionality for summarising model statistics and the output. We finally get the posterior distribution for \BLOS\ and $\sigma$. One particular statistic of interest of these probability distributions is the Highest Posterior Density (HPD) interval. A 100 (1 - $\alpha$)\% HPD interval is a region that satisfies the following two conditions:

\begin{itemize}
\item The posterior probability of that region is 100(1 - $\alpha$)\%. 

\item The minimum density of any value within that region is equal to or larger than the density of any point outside that region.
\end{itemize}

An example of the resulting posterior distributions for one pixel is shown in Fig.~\ref{Figure:6}. In order to summarise the posterior distribution of each pixel, the mode of $B_{\mathrm{LOS}}$  will be given as its representative value, combined with the 95\% HPD interval as a measure of the most probable values of $B_{\mathrm{LOS}}$.

\section{Results} \label{results}

In this section we present the results of the Bayesian inference. The association of the measured line-of-sight magnetic field components with spicules is done in Sect.~\ref{sec:association}.

\subsection{Off-limb pixels}\label{sec:off-limb}

The method described in Sect.~\ref{sec:Data} is applied to all pixels located above the limb in each data set, where the limb is defined as the sharp intensity transition in the wings of the Stokes~$I$ profile. The top panel of Fig.~\ref{Figure:7} shows spicules from data set \#1 with the points that have passed the correlation and asymmetry criteria for a particular time instant plotted on the corresponding intensity image. The majority of these points are located between 2\arcsec\ and 5\arcsec\ (1450 and 3625 km) above the limb, the reason being that both the pixels at a height smaller than 2~Mm, in which superposition becomes too strong for the WFA to be applicable (see Sect.~\ref{sec:superposition}), and the pixels of faint spicular material at larger heights fail to meet the two criteria.

We note that data set \#1 is the only one that contains off-limb spicules far from an active region. The lack of a predominant north or south magnetic polarity in the set of spicules coupled with their wide range of inclinations allows for both positive and negative $B_{\mathrm{LOS}}$, as evidenced by the values in the colour bar of Fig.~\ref{Figure:7}, top. In addition, this figure also indicates that, although not many pixels pass our two selection criteria, $|B_{\mathrm{LOS}}|$ values well in excess of 100 G prevail. In Fig.~\ref{Figure:11} all the inferred values of $B_{\mathrm{LOS}}$ for each of the off-limb data sets listed in Table~\ref{table1} are put together by means of a histogram. The top left panel, corresponding to data set \#1, reveals that the number of pixels for which the LOS magnetic field could be determined is considerable and that |$B_{\mathrm{LOS}}$| reaches values as high as 800 G. A summary of the results of Fig.~\ref{Figure:11} is shown in Table~\ref{table:2}, in which the mode and the 95\% HPD interval of the distributions shown in Fig.~\ref{Figure:11} are computed, using the absolute value of $B_{\mathrm{LOS}}$.

Next we present the results of data set \#8, an example of an active region area with off-limb spicules. The bottom panel of Fig.~\ref{Figure:7} is analogous to the top one in that line-of-sight magnetic field components greater than 100~G are abundant in the region between 2\arcsec\ and 5\arcsec\ above the limb and that WFA inversions are absent outside this area. The main differences between the two figures are that it has been possible to carry out the Bayesian inversion for many more pixels and that all measured $B_{\mathrm{LOS}}$ are positive. This is also true for the whole duration of this data set, such as shown by the histogram in Fig.~\ref{Figure:11}  (second row, centre). This result is very robust because the number of pixels of data set~\#8 whose $B_{\mathrm{LOS}}$ has been computed is about an order of magnitude larger than that of data set \#1. Hence, we have found strong evidence of magnetic fields that point toward the observer whose LOS components are clustered around 200~G; in particular, the mode of their distribution is 193~G (see Table~\ref{table:2}). But what is the reason for this particular orientation of the magnetic field vectors in the whole area? To explain this we make use of Fig.~\ref{Figure:10}, where a sketch of the region observed in data sets \#2, \#4, \#6 and \#8 is shown. NOAA AR 12551 had a leading south magnetic polarity and a trailing north magnetic polarity that at the observing time of data set~\#8 were both on the non-visible side of the west limb. Assuming that the magnetic field of spicules in the close vicinity of each active region magnetic polarity is not randomly oriented but has the same orientation of this polarity \citep[such as hypothesised by][]{orozco2015}, all spicules visible in data set~\#8 possess a magnetic field pointing toward the observer, except for a tiny number of pixels with negative $B_{\mathrm{LOS}}$ around $-$100~G (see Fig.~\ref{Figure:11}, second row, centre), that do not follow this orientation pattern.

Now we turn our attention to data sets \#2, \#4 and \#6, that were acquired a few hours before data set \#8 in a time span of 90 min. Figure~\ref{Figure:10} suggests that at the time when these three data sets were taken, the LOS intersected a north polarity spicule area with the magnetic field pointing away from the observer, hence the negative sign of $B_{\mathrm{LOS}}$ in the histograms of Fig.~\ref{Figure:11}. Once more, |$B_{\mathrm{LOS}}$| is greater than 200~G for a considerable number of pixels (see Table \ref{table:2} for the |$B_{\mathrm{LOS}}$| mode of each data set) and a minority of pixels with a polarity opposite to the prevailing one are also present.

Data set \#10 was acquired above an active region on the east limb. Figure~\ref{Figure:11} (bottom right panel) evidences that this is the case in which we find the highest number of pixels that have passed the filters for the computation of $B_{\mathrm{LOS}}$. The spicules in this data set are near the north magnetic polarity of the active region that, at the observing time, had not reached the limb yet. Therefore, the magnetic field in the spicules points towards the observer and positive $B_{\mathrm{LOS}}$ are obtained. This result reinforces our previous hypothesis of a dominant spicule polarity in the areas adjacent to the dominant active region polarities. In this case the $B_{\mathrm{LOS}}$ cluster around a much higher value than for data sets \#2, \#4, \#6 and \#8, namely around 450 G (Table \ref{table:2}).

In this section we have presented the results for off-limb pixels and, in order to explain the prevailing positive or negative signs of the inferred \BLOS\ in all data sets except \#1, have put forward the hypothesis that this is the consequence of spicules near a sunspot having the magnetic polarity of that sunspot. In Sect.~\ref{sec:association} we provide evidence that the pixels for which the application of the WFA has given a \BLOS\ value actually belong to spicules. Furthermore, an inspection of the distributions of Fig.~\ref{Figure:11} shows that the largest \BLOS\ values correspond to both QS spicules (data set \#1) and spicules near an AR (data set \#10). Hence, there seems to be no difference in the magnetic field strength of spicules in the QS and near an AR. Moreover, we mentioned before that the number of pixels for which \BLOS\ could determined is much smaller for data set \#1 than for the other data sets. Although this might be caused by the worse seeing of data set \#1, we offer another explanation. The superposition along the LOS of different structures may lead to observed Stokes $I$ and $V$ that do not pass the imposed criteria before the WFA is applied (see Sect.~\ref{sec:superposition}). If the magnetic field orientation of spicules near an AR is more constrained than that of their QS counterparts, as we have suggested above, then QS pixels have a higher probability of containing overlapping spicules with both positive and negative \BLOS, which will result in a smaller number of pixels for which \BLOS\ can be obtained.

\begin{table}
\caption{Off-limb pixels. The second and third columns give the mode and 95\% HPD interval of the absolute value of the $B_{\mathrm{LOS}}$ distributions shown in Fig.~\ref{Figure:11}. Values are given in G.}             
\label{table:2}      
\centering                          
\begin{tabular}{|c|c|c|}        
\hline\hline                 
Data set& |$B_{\mathrm{LOS}}$|& 95\% HPD\\    
\hline                        
   1 & 187 & [73, 742]  \\\hline        
   2 & 197& [92, 408]\\\hline  
   4 &211  & [111, 461]  \\\hline  
   6 &196 & [112, 385]  \\\hline  
    8&193 & [98, 384] \\ \hline  
   10&450 & [219, 647] \\ 
\hline                                   
\end{tabular}
\end{table}

\subsection{On-disk pixels} \label{sec:on-disk}

The same analysis is done for the pixels seen on the solar disk, which are not in the vicinity of an active region and hence are not expected to possess a predominant north or south magnetic polarity. Signal superposition (see Sect.~\ref{sec:superposition}) strongly limits the number of pixels that satisfy the correlation and asymmetry conditions. This can be seen in Fig.~\ref{Figure:12}, where similar histograms to those of Fig.~\ref{Figure:11} are shown for the data sets that correspond to disk pixels. As predicted for QS pixels, both signs of $B_{\mathrm{LOS}}$ are inferred. Table~\ref{table:3} provides a summary of the results for the on-disk spicules in a similar manner as Table \ref{table:2}. Based on the results of Sect.~\ref{sec:superposition}, we suggest that smaller $B_{\mathrm{LOS}}$ values are obtained on the disk because of the stronger influence of superposition of the absorbing plasma elements on top of the chromosphere.

\begin{table}
\caption{On-disk pixels. The second and third columns give the mode and 95\% HPD interval of the absolute value of the $B_{\mathrm{LOS}}$ distributions shown in Fig.~\ref{Figure:12}. Values are given in G.}             
\label{table:3}      
\centering                          
\begin{tabular}{|c|c|c|}        
\hline\hline                 
Data set& |$B_{\mathrm{LOS}}$|& 95\% HPD\\    
\hline                        
   3 & 63& [92, 408]  \\\hline        
   5 & 86& [44, 168] \\\hline  
   7 &100  & [12, 150]  \\\hline  
   9 & 83& [33, 132]  \\\hline  
    11& 148& [91, 212]  \\ \hline  
\end{tabular}
\end{table}

\begin{figure*}
   \centering
   \includegraphics[width=8cm]{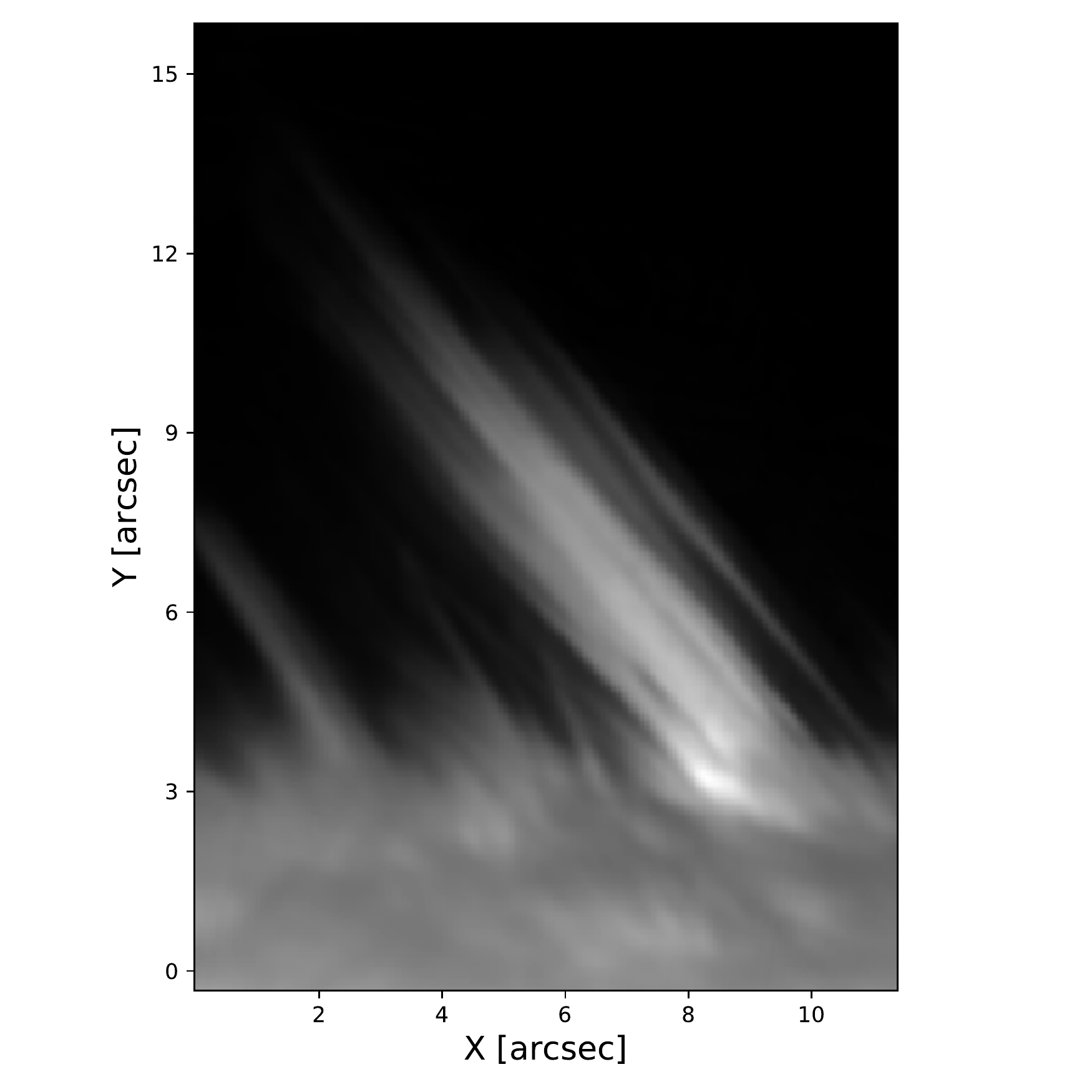}
   \includegraphics[width=8cm]{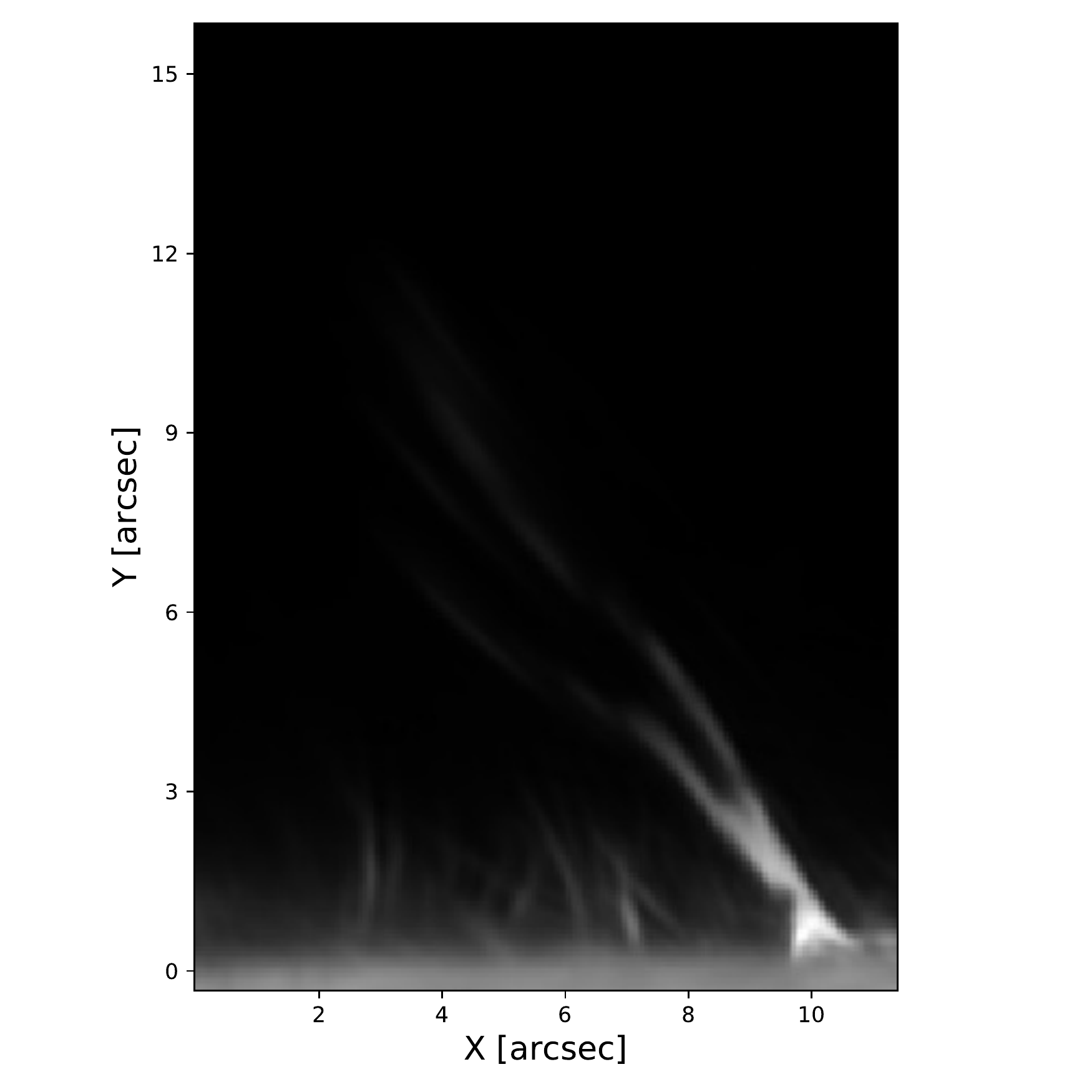}
   \caption{Chromospheric anemone observed on 03 June 2016 at 16:31:50 UT, viewed in the \ion{Ca}{II} 8542 \AA\ line centre (left) and $-0.75$ \AA\ from line centre (right).}
              \label{Figure:13}%
    \end{figure*}

\begin{figure*}
   \centering
   \includegraphics[width=15cm]{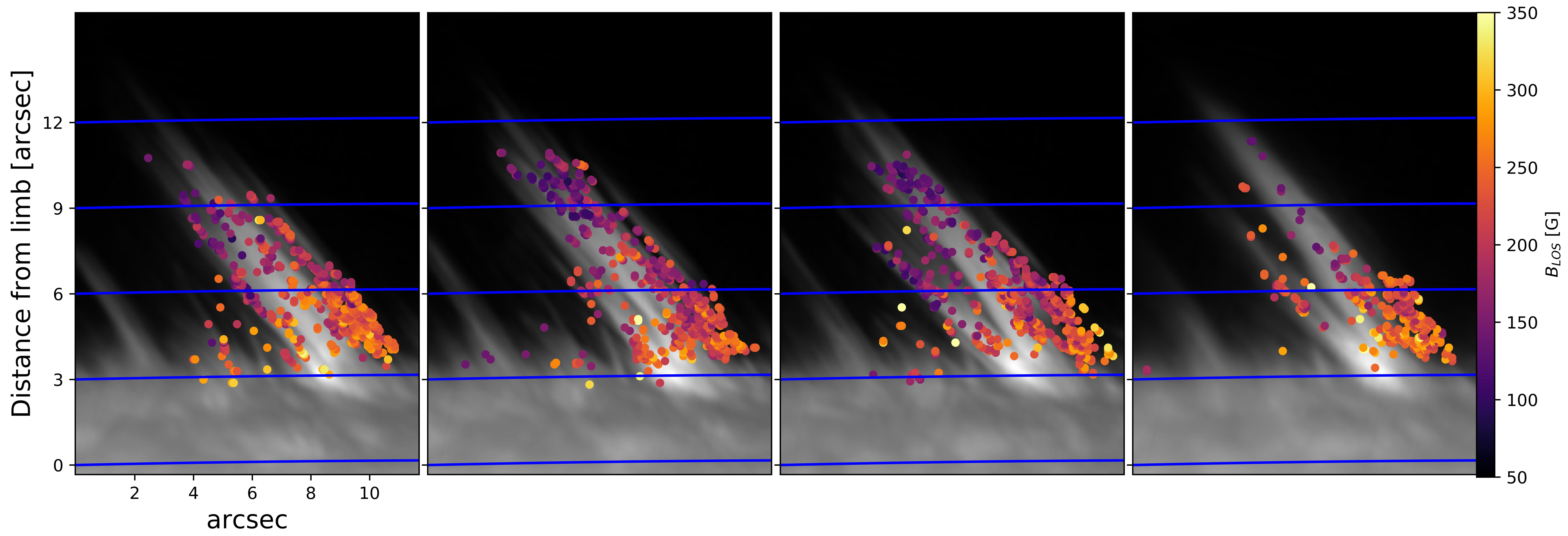}
   \caption{Temporal evolution of the chromospheric anemone seen in Fig.~\ref{Figure:13} and the inferred \BLOS\ values. The leftmost picture was taken on 03~June 2016 at 16:31:50 UT and the subsequent figures show the evolution of the region and the inferred \BLOS\ at time intervals of 36.33 s.}
              \label{Figure:14}%
    \end{figure*}

\subsection{Chromospheric anemone} \label{anemonesect}

The structure near coordinates (10\arcsec, 35\arcsec) of Fig.~\ref{Figure:2} is a chromospheric anemone \citep{Shibata1591}. It is a type of chromospheric jet which possesses an inverted Y-shape structure (see Fig.~\ref{Figure:13}) attributed to magnetic reconnection. Even though this work is focused on spicules, the WFA can also be applied to this structure, as shown in Fig.~\ref{Figure:14}. The isolation and higher \ion{Ca}{II} 8542 \AA\ line intensity of the anemone provide clearer $I$ and $V$ signals, something difficult to obtain in the lower lying spicules, which frequently overlap on top of each other. A comparison  between Figs. \ref{Figure:7} and \ref{Figure:14} shows that the line-of-sight magnetic field component in the anemone and the neighbouring spicules takes similar values. There is also an apparent tendency of $B_{\mathrm{LOS}}$ to decrease with height along the anemone, an issue that will be discussed in Sect.~\ref{Sect:52}. The distribution of $B_{\mathrm{LOS}}$ in the anemone is analogous to that shown in Fig.~\ref{Figure:11}. The mode of the distribution is 210 G and its 95\% HPD interval is [111, 293]~G.

\section{Discussion} \label{sec:discussion}

\subsection{Signal superposition and integration time} \label{sec:superposition}
We now study the effect that structures above the limb overlapping on the same detector pixel have on the \BLOS\ obtained from the WFA. The plan is to specify the Stokes parameters of two such structures that lie along the same LOS, construct from them the observed Stokes parameters and use Eq.~(\ref{eq:3}) to calculate \BLOS. For this task we use the toy model of \citet{Centeno2010}, made of two slabs illuminated from below and whose radiation is detected by an observer; see their Fig.~4. The radiation emitted by slab~1, the one furthest from the observer, must cross slab~2 before reaching the detector. \citet{Centeno2010} model this effect by a wavelength-independent optical depth, $\tau_2$. We consider an optically thin plasma and hence use $\tau_2\ll 1$. The expressions for the detected $I$ and $V$ profiles, namely Eqs.~(2) and~(3) of \citet{Centeno2010}, reduce to:

\begin{equation}\label{eq:Iobs}
I_\mathrm{obs} \simeq I_1 + I_2,
\end{equation}

\begin{equation}\label{eq:Vobs}
V_\mathrm{obs} \simeq V_1 + V_2,
\end{equation}

\noindent where $I_i$, $V_i$ ($i=1,2$) are the Stokes $I$ and $V$ emitted by the slabs. In the case of slab 2 being optically thick, $V_\mathrm{obs} \simeq V_2$ and $I_\mathrm{obs} \simeq I_2$, and therefore  the measurements of \BLOS\ will be dominated by the magnetic field of slab 2.

To construct the observed Stokes parameters with Eqs.~(\ref{eq:Iobs}) and (\ref{eq:Vobs}) we must specify $I$ and $V$ of the two slabs. Their intensity profiles are chosen as Gaussians:

\begin{equation}\label{eq:I1I2}
I_1 = I_{01}\exp\left[-\frac{(\lambda-\lambda_\mathrm{c1})^2}{\sigma_\mathrm{I}^2}\right], \hspace{3ex} I_2 = I_{02}\exp\left[-\frac{(\lambda-\lambda_\mathrm{c2})^2}{\sigma_\mathrm{I}^2}\right],
\end{equation}

\noindent where we set $\sigma_\mathrm{I}=0.5$~\AA, which is the typical width of the off-limb intensity profiles in our data sets. Moreover, $\lambda_\mathrm{ci}$, $i=1,2$, is the Doppler shifted line centre wavelength, that is, $\lambda_\mathrm{ci}=\lambda_0(1+v_\mathrm{Di}/c)$ with $\lambda_0=8542.1$~\AA, and $v_\mathrm{Di}$ is the LOS velocity of slab~$i$. Eq.~(\ref{eq:3}) with $f=1$ is used to calculate $V_{1,2}$ from $I_{1,2}$ given the LOS magnetic field components in the two slabs, $B_\mathrm{LOS1,2}$.

Now, the Stokes profiles of each slab are characterised by the parameters $I_{0i}$, $v_\mathrm{Di}$ and $B_\mathrm{LOSi}$, which makes a total of six parameters. We recall that when the LOS component of $\vec{B}$ points toward (away from) the observer, then \BLOS\ is positive (negative). One of the intensity amplitudes, $I_{01}$ and $I_{02}$, can be set to an arbitrary value and the results do not change as long as the ratio $I_{02}/I_{01}$ is not altered. We thus choose $I_{01}=1$ and leave $I_{02}/I_{01}$ free. In addition, the \BLOS\ derived from the WFA inversion of $I_\mathrm{obs}$ and $V_\mathrm{obs}$ remains almost unchanged when the Doppler velocities are varied in such a way that the difference $v_\mathrm{D2}-v_\mathrm{D1}$ is kept constant. Our problem then contains four free parameters: $B_\mathrm{LOS1}$, $B_\mathrm{LOS2}$, $I_{02}/I_{01}$ and $\Delta v_\mathrm{D}\equiv v_\mathrm{D2}-v_\mathrm{D1}$. Once they are fixed, the Stokes $I$ and $V$ of each slab can be determined with Eqs.~(\ref{eq:I1I2}) and (\ref{eq:3}) and their superposition can be derived from Eqs.~(\ref{eq:Iobs}) and (\ref{eq:Vobs}). The final step is to fit a straight line to $V_\mathrm{obs}$ vs $\partial I_\mathrm{obs}/\partial\lambda$ and to use Eq.~(\ref{eq:BLOS_slope}) to compute the resulting \BLOS. To accept the result of this fit we impose the restriction $|R|>0.9$, where $R$ is the Pearson correlation coefficient.

\begin{figure}[h!]
  \centering
  \includegraphics[width=8cm]{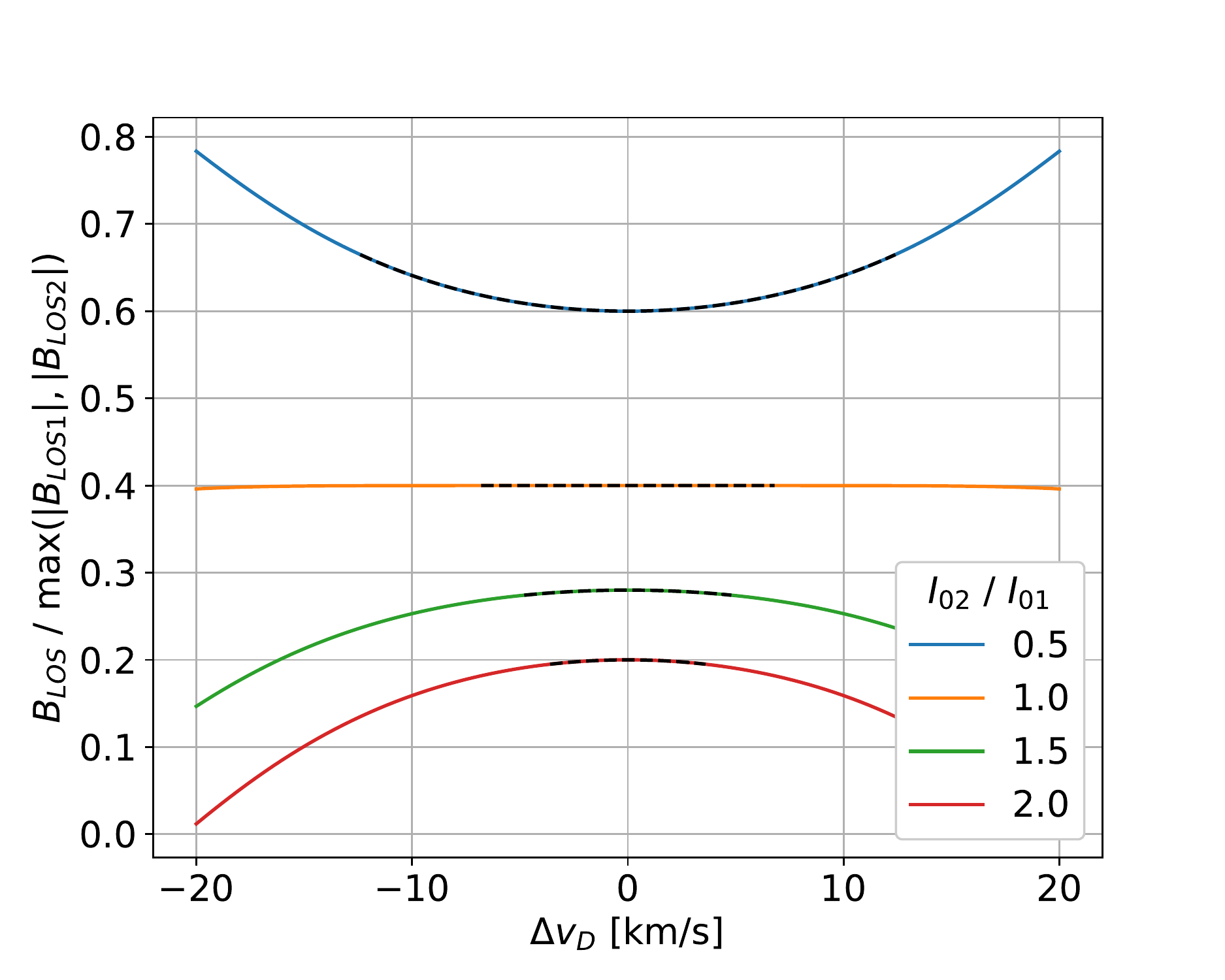}
  \includegraphics[width=8cm]{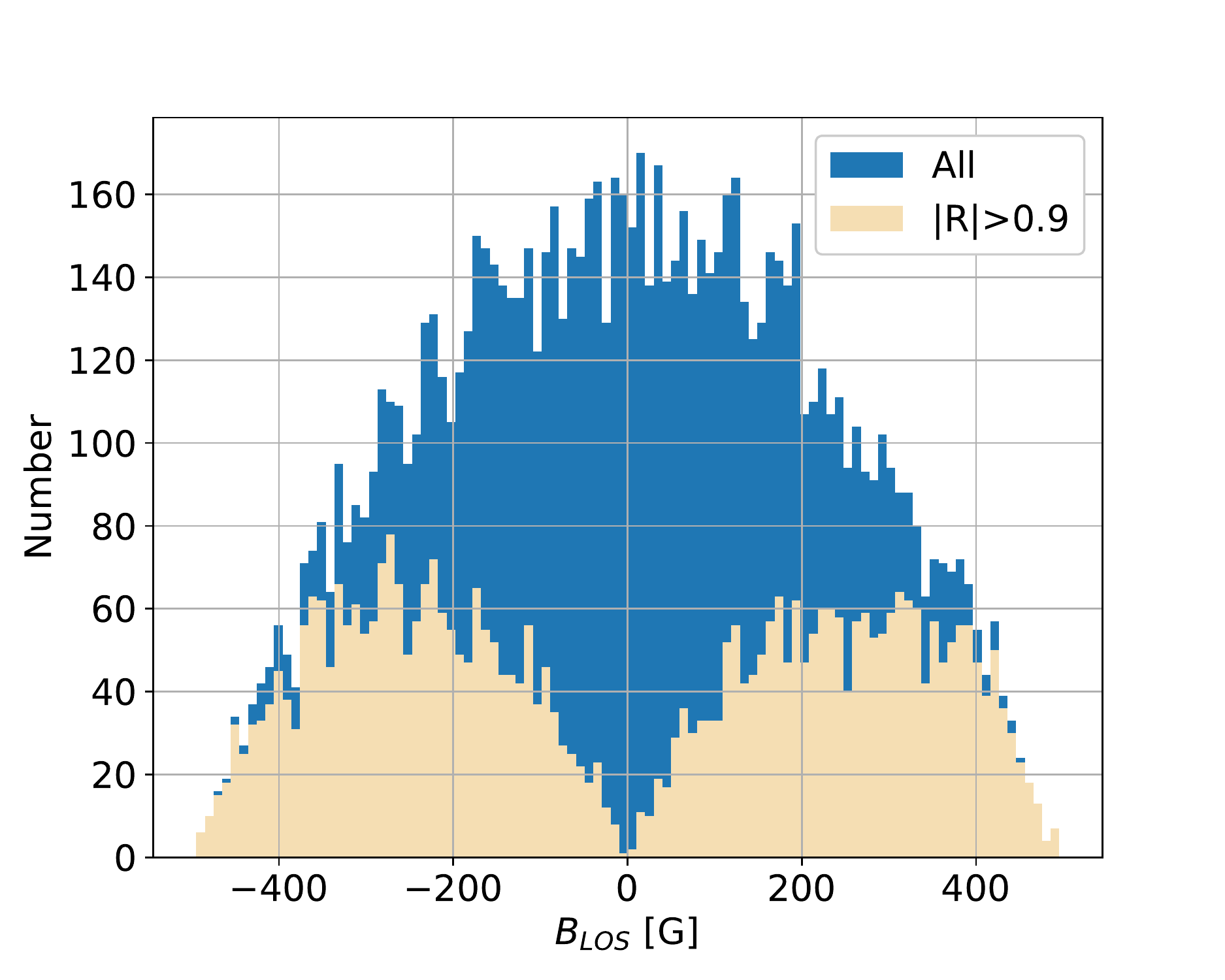}
  \includegraphics[width=8cm]{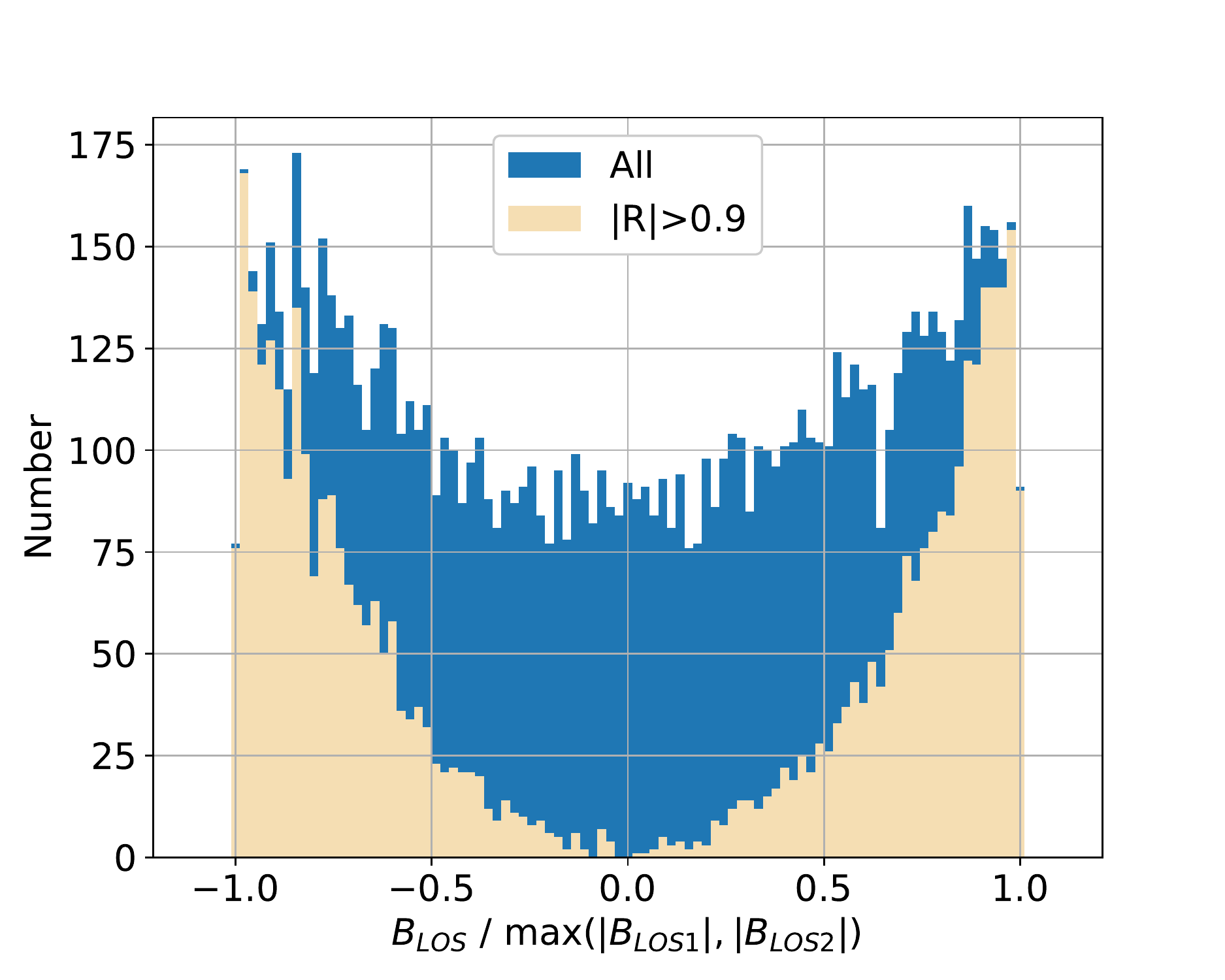}
  \caption{Results of the WFA \BLOS\ inversion when the Stokes $I$ and $V$ parameters in a pixel come from two overlapping structures (cf. Eqs.~[\ref{eq:Iobs}] and [\ref{eq:Vobs}]). Top: \BLOS\ normalised by dividing it by the maximum LOS magnetic field of the two structures, in absolute value. It is plotted vs the difference between the two slabs' Doppler velocity, $\Delta v_\mathrm{D}$, for several intensity amplitude ratios, $I_{02}/I_{01}$. The dashed black lines give the results for which $|R|>0.9$. In this plot $B_\mathrm{LOS1}=200$~G and $B_\mathrm{LOS2}=-40$~G. Middle: \BLOS\ distribution from the Monte Carlo experiment. Bottom: \BLOS\ distribution divided by the maximum of $|B_\mathrm{LOS1}|, |B_\mathrm{LOS2}|$. In the middle and bottom panels, the blue and beige colours respectively correspond to all the samples in the Monte Carlo experiment and only those that pass the filter $|R|>0.9$.}
  \label{fig:superposition}
\end{figure}

The results presented here have been computed with a set of 15 wavelengths uniformly distributed in the range [$\lambda_0-1.5$~\AA, $\lambda_0+1.5$~\AA]. Using the non-uniformly spaced wavelengths of our data sets does not substantially modify the results. When both Doppler shifts are equal ($\Delta v_\mathrm{D}=0$) and for equal intensity amplitudes in both slabs ($I_{02}/I_{01}=1$), but for any value of $B_\mathrm{LOSi}$, we find that the \BLOS\ given by the WFA is equal to the average of the slabs' LOS magnetic field components, $(B_\mathrm{LOS1}+B_\mathrm{LOS2})/2$, with a correlation coefficient $|R|=1$. To describe the effect of signal superposition for $\Delta v_\mathrm{D}\neq 0$ and/or $I_{02}/I_{01}\neq 1$ we consider the specific values $B_\mathrm{LOS1}=200$~G and $B_\mathrm{LOS2}=-40$~G. The top panel of Fig.~\ref{fig:superposition} presents the obtained \BLOS\ as a function of $\Delta v_\mathrm{D}$ and $I_{02}/I_{01}$. To allow a comparison with the maximum LOS magnetic field component present in both slabs, the \BLOS\ returned by the WFA is normalised to the maximum of $|B_\mathrm{LOS1}|, |B_\mathrm{LOS2}|$. In addition, the dashed black lines indicate the parameter values for which the condition $|R|>0.9$ is met. Thus, the top panel of Fig.~\ref{fig:superposition} shows that the \BLOS\ from the WFA inversion is between 20\% (red curve) and 67\% (left and right ends of black line superimposed on blue curve) the maximum \BLOS\ in the two slabs. These limits vary for other choices of $B_\mathrm{LOS1}$, $B_\mathrm{LOS2}$ but the upper one is always smaller than or equal to 100\% and only when $B_\mathrm{LOS1}\simeq B_\mathrm{LOS2}$ does it approach 100\% (from below). Hence, we conclude that the application of the WFA to the Stokes signals from two overlapping structures leads to a reduction of the largest \BLOS\ present in both of them.

All curves in the top panel of Fig.~\ref{fig:superposition} show that large Doppler velocity differences between the two slabs spoil the WFA inversion. These curves are symmetric about $\Delta v_\mathrm{D}=0$, as expected. The largest values of $|\Delta v_\mathrm{D}|$ for which the WFA works are found for the blue curve, which is the one for which the slab with the largest \BLOS\ also has the largest $I$ amplitude (slab 1 in this example). The blue curve also gives the largest \BLOS\ from the WFA inversion. Therefore, the combination of large \BLOS\ and large intensity amplitude in the same slab favour the validity of the WFA formula and lead to inferred \BLOS\ values closer to that of this slab.

To see the effect of signal superposition on a large set of pairs of overlapping structures we carry out a Monte Carlo (MC) experiment. We randomly set 10,000 slab pairs with their $B_\mathrm{LOS1}$ and $B_\mathrm{LOS2}$ uniformly distributed between $-$500~G and 500~G, $v_\mathrm{D1}$ and $v_\mathrm{D2}$ uniformly distributed between $-40$~km~s$^{-1}$ and 40~km~s$^{-1}$ and $I_{02}/I_{01}$ uniformly distributed between 0.5 and~2. The results are presented in the middle and bottom panels of Fig.~\ref{fig:superposition} as histograms of the \BLOS\ calculated from the WFA inversion; the blue and beige colours correspond to all the MC samples and to only those for which $|R|>0.9$, respectively. The middle panel shows that, although the probability of near zero \BLOS\ is very large when all MC samples are considered, imposing $|R|>0.9$ removes most results with small $|B_\mathrm{LOS}|$. This can partly explain the scarcity of $|\BLOS|\lesssim 50$~G in the histograms of data sets~\#1 (Fig.~\ref{Figure:11}) and \#3, \#5, \#9, \#11 (Fig.~\ref{Figure:12}). The middle panel also shows that intermediate \BLOS\ values centered about $\pm 300$~G become the most common. This is a consequence of the random selection of $B_\mathrm{LOS1}$ and $B_\mathrm{LOS2}$. When presenting the \BLOS\ histograms of data sets~\#1, \#3, \#5, \#9 and \#11 we suggested that they contained negative and positive LOS magnetic field components because the spicules had no preferred orientation. The middle panel of Fig.~\ref{fig:superposition} has been derived with a random orientation of the magnetic field vector and therefore it is not surprising that its overall \BLOS\ distribution resembles that of data sets~\#1, \#3, \#5, \#9 and \#11.

By representing the histogram of \BLOS\ normalised to the maximum of $|B_\mathrm{LOS1}|, |B_\mathrm{LOS2}|$ (Fig.~\ref{fig:superposition}, bottom), we see that in many samples the strongest \BLOS\ of the two slabs is retrieved after applying the WFA to the detected Stokes signals. Smaller \BLOS\ can also be obtained, although with a decreasing probability as we approach $|B_{\mathrm{LOS}}|=0$. Finally, the occurrence of $|B_{\mathrm{LOS}}|\simeq 0$ is negligible. The conclusion is that signal superposition of two slabs cannot produce a \BLOS\ determined with the WFA that is larger than the strongest \BLOS\ in both slabs. Moreover, for randomly oriented magnetic fields, Doppler velocities and $I$ amplitudes in the two slabs, the WFA has a large probability to return a \BLOS\ close to the strongest one in both slabs, although much smaller values cannot be discarded.

In the middle and bottom panels of Fig.~\ref{fig:superposition}, almost 60\% of the MC samples do not meet the $|R|>0.9$ criterion. If the Doppler velocity limits are reduced from $\pm 40$~km~s$^{-1}$ to $\pm 20$~km~s$^{-1}$, still half the samples do not satisfy $|R|>0.9$. This means that in our highly idealised model with only two structures, signal superposition can ``degrade'' the Stokes parameters so as to make them useless for the WFA inversion under our constraints. We can hint that Eq.~(\ref{eq:3}) applied to Stokes signals coming from more than two overlapping off-limb structures will behave in a similar manner and that the degradation of the Stokes parameters will be such that less pixels will have Stokes parameters useful for the WFA inversion. And those pixels in which the WFA can be applied will have a larger probability of returning a \BLOS\ smaller than the largest LOS magnetic field component involved.

Furthermore, a long integration time can lead to a structure's emission not being recorded all the time in the detector pixel if it appears, disappears or moves sideways; or to its emission changing in time; or to the presence of overlapping structures with time dependent \BLOS, $v_\mathrm{D}$ and $I_0$, etc. None of these effects have been included in the toy model, but based on the results of this section we suggest that either the resulting Stokes $I$ and $V$ parameters will not satisfy the correlation and asymmetry criteria or, if they do, that the application of the WFA will yield a \BLOS\ that is not larger than any LOS magnetic field component along the line-of-sight during the whole integration time.

Regarding on-disk spicules, they can also overlap along the LOS with other spicules, but, above all, their Stokes parameters are ``contaminated'' by the underlying chromosphere, which is highly dynamic and inhomogeneous. This superposition introduces a wider range of variation of the parameters \BLOS, $I_0$, $v_\mathrm{D}$,~\ldots Hence, we speculate that the problems we have described for two or more off-limb overlapping structures will become worse and that much less on-disk than off-limb pixels will have observed Stokes $I$ and $V$ that pass the correlation and asymmetry criteria. This can explain the difference in the number of pixels in the histograms of Figs.~\ref{Figure:11} and~\ref{Figure:12}.

\begin{figure*}
   \centering
   \includegraphics[width=15cm]{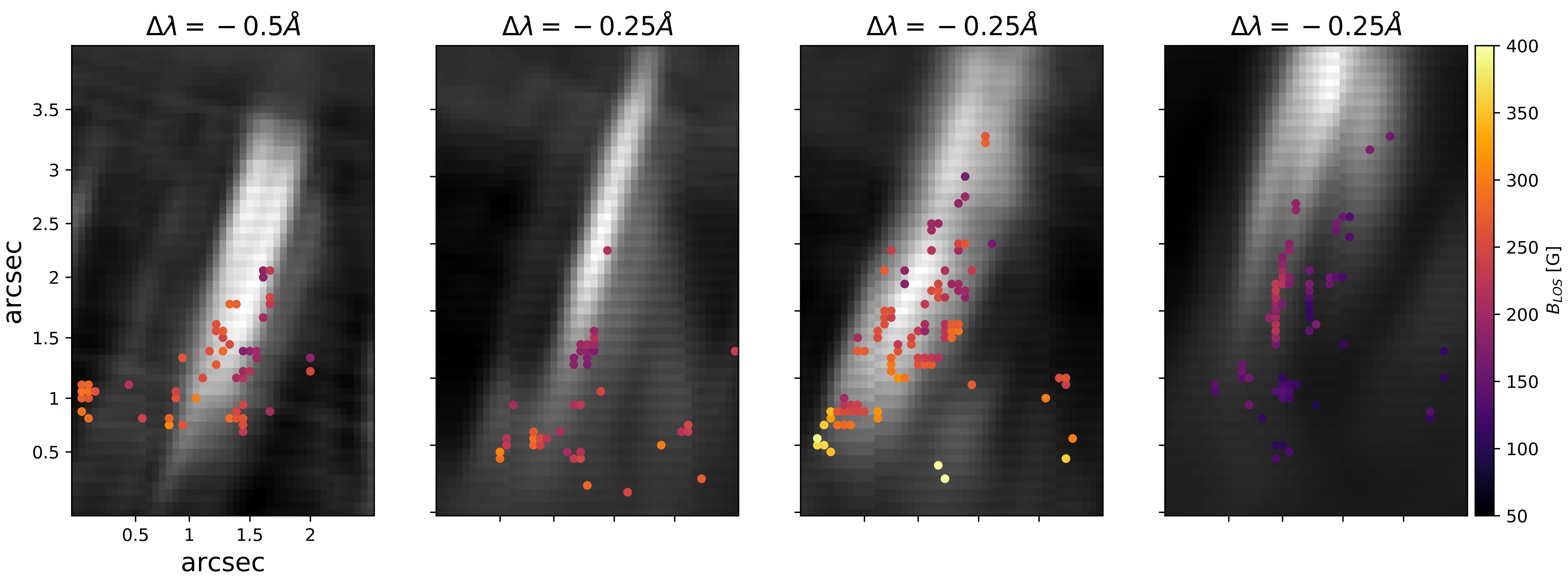}
   \caption{Line-of-sight magnetic field results of the red and blue rectangles of the bottom panel of Fig.~\ref{Figure:7} are overplotted on top of Stokes $I$ at different times: 16:42:44~UT, 16:50:36~UT, 16:57:36~UT, 17:03:55~UT. The first three panels correspond to the red rectangle and the last panel to the blue rectangle. The difference, $\Delta\lambda$, from line centre is shown at the top of each panel. A radial filter \citep[see][]{skogsrud2015} has been applied to the images to enhance spicular structures.}
              \label{Figure:9}%
    \end{figure*}    

\subsection{Association of measured \BLOS\ with spicules} \label{sec:association}


To prove the association of our off-limb \BLOS\ measurements with spicules, we consider the red and blue rectangles of the bottom panel of Fig.~\ref{Figure:7} and plot the Stokes~$I$ parameter in these rectangles and the associated \BLOS\ for several times. The results are displayed in Fig.~\ref{Figure:9}, that confirms that each pixel for which the Bayesian inversion has been carried out lies on a spicule.

\begin{figure*}
   \centering
   \includegraphics[width=8cm]{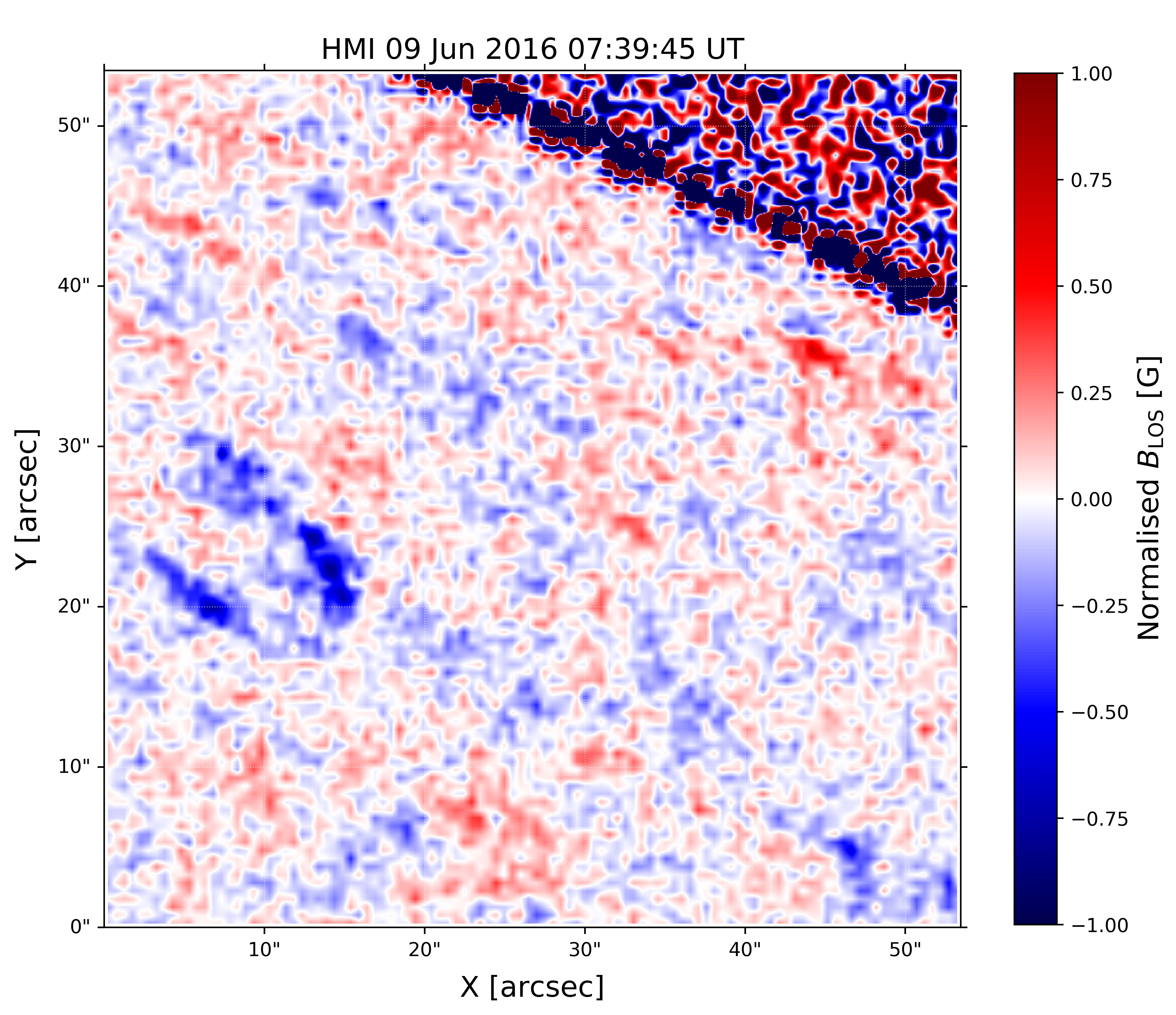}
   \includegraphics[width=8cm]{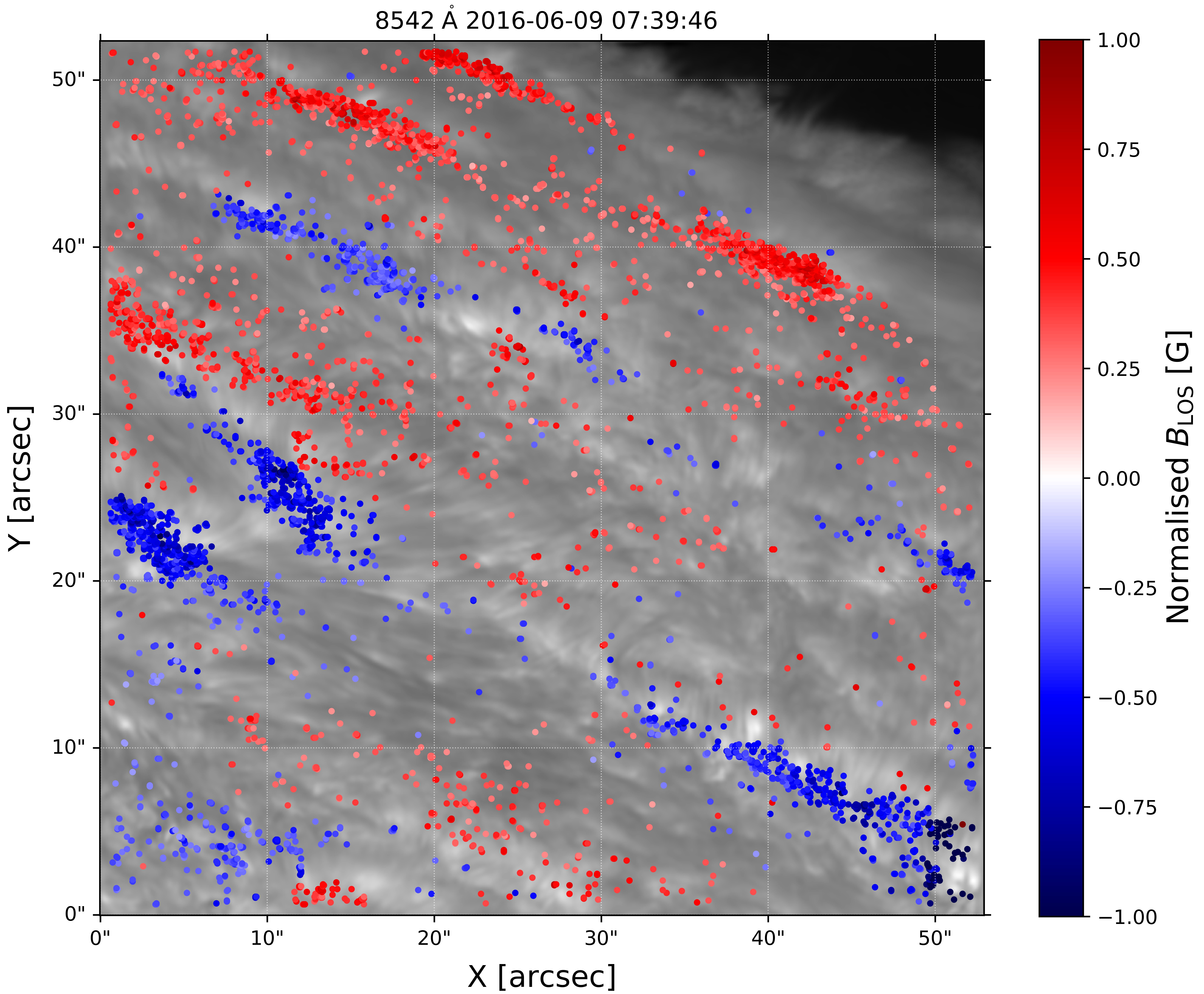}
   \caption{Left: photospheric $B_{\mathrm{LOS}}$ obtained with SDO/HMI. Right: $B_{\mathrm{LOS}}$ obtained in this work. In each panel the $B_{\mathrm{LOS}}$ values have been normalised to their largest respective value in the whole field-of-view. The observing time is given on top of each image.}
              \label{Figure:15}%
    \end{figure*}

\begin{figure*}
   \centering
   \includegraphics[width=8cm]{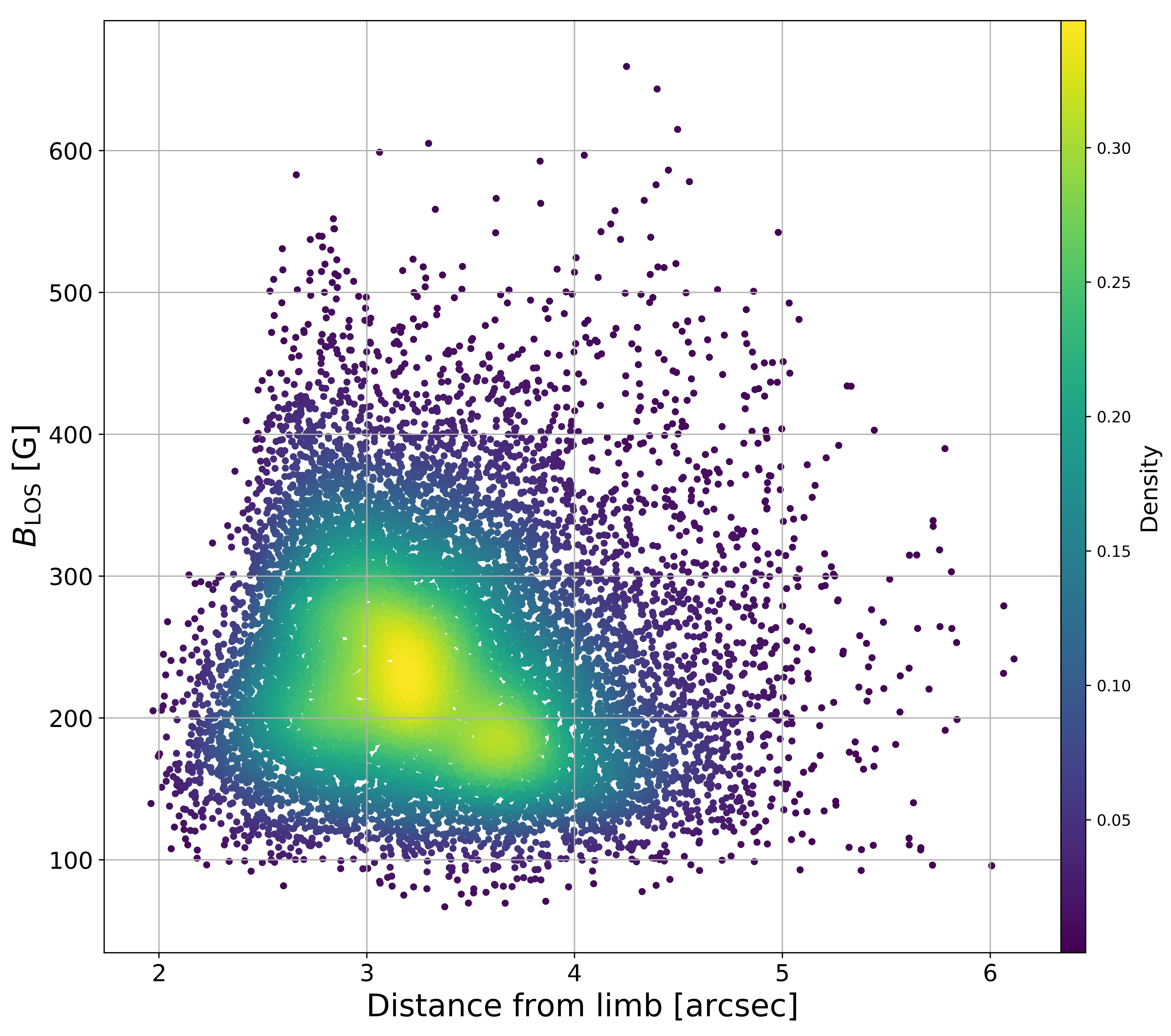}
   \includegraphics[width=8cm]{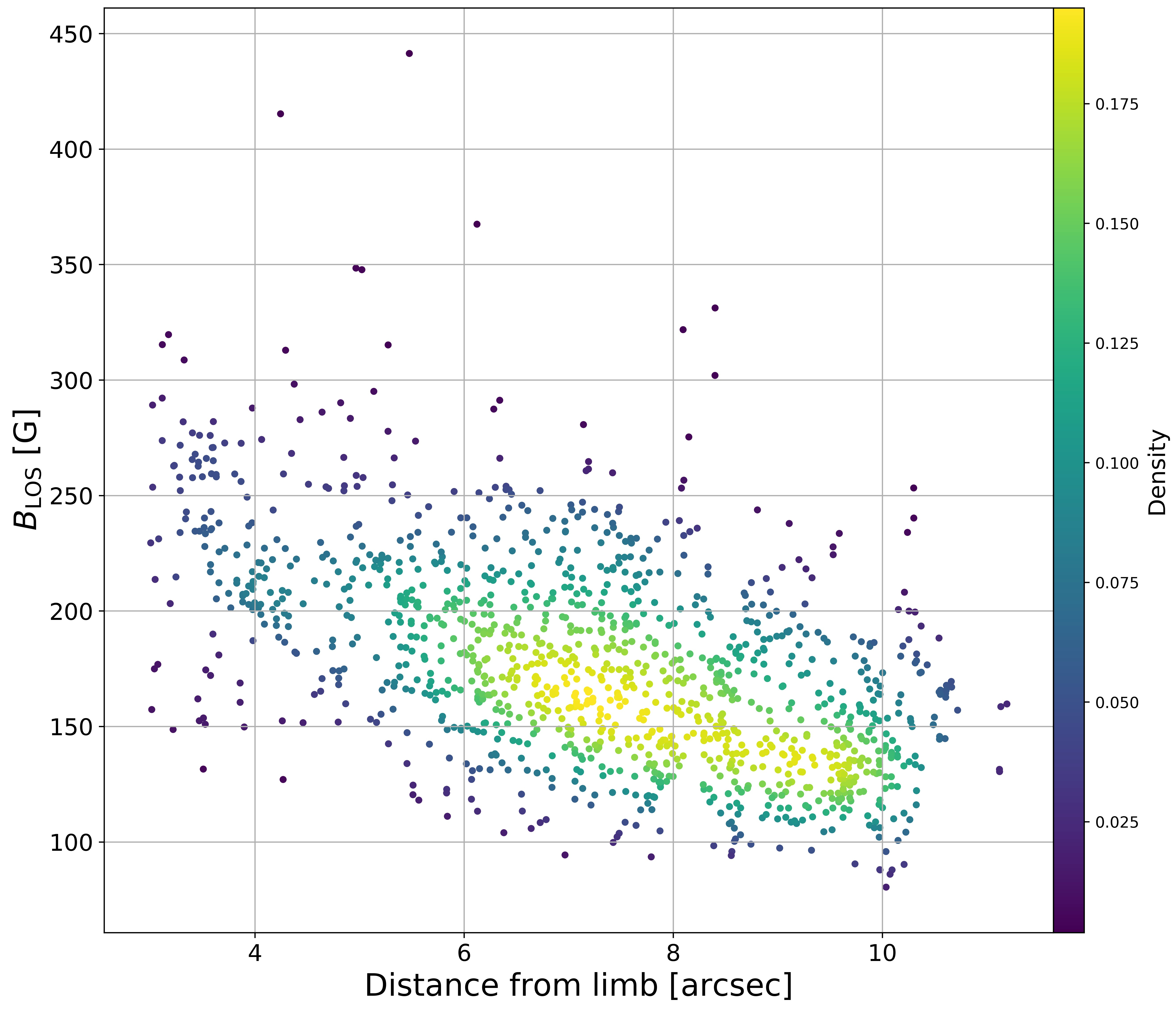}
   \caption{Left: $B_{\mathrm{LOS}}$ vs distance from the limb for all pixels in data set \#8. Right: Same plot for the chromospheric anemone discussed in Sect. \ref{anemonesect}. The colours in the plot represent the local density of points.}
              \label{Figure:16}%
    \end{figure*}    



The identification of on-disk \BLOS\ measurements with spicules is more problematic because both the spicules and the underlying chromosphere can contribute to the line formation. Here we compare the spatial distribution of the pixels for which the Bayesian inversion has been performed with photospheric LOS magnetic field data taken with the Helioseismic and Magnetic Imager \citep[HMI;][]{Scherrer2012,Schou2012} instrument on board SDO. HMI's hmi.B\_45s data products provide the line-of-sight component of the photospheric magnetic field from observations in the \ion{Fe}{I} 6173.3 \AA\ line every 45 s with the HMI Doppler camera. The pixel size of HMI's camera is 0.5\arcsec\ and for this reason the original image data have been interpolated in order to match CRISP's pixel scale. A comparison between the results of the two instruments is shown in Fig.~\ref{Figure:15}. The values of $B_{\mathrm{LOS}}$ are normalised to the maximum value in each data set in order to compare the spatial distribution of the sign of the $B_{\mathrm{LOS}}$ values. Only a few on-disk pixels meet the asymmetry and correlation criteria described in Sect.~\ref{sec:Data}, rendering the comparison between the results difficult. For this reason, to produce the right panel of Fig.~\ref{Figure:15} the correlation criterion has been lowered, accepting pixels where |$R$| > 0.6 (while maintaining $\varepsilon=0.5$ for the asymmetry criterion). As Fig.~\ref{Figure:15} shows, there is a strong spatial correlation in the sign distribution of $B_{\mathrm{LOS}}$ in the photosphere and in the lower chromosphere. Given that the photospheric magnetic fields in this image delineate the chromospheric network, from which spicules protrude, we have a strong indication that our on-disk measurements give the $B_{\mathrm{LOS}}$ of spicules. However, we cannot rule out that a portion of these results come from structures other than spicules.

\subsection{Height variation of \BLOS} \label{Sect:52}

The expansion of the magnetic field with increasing height leads to lower magnetic field values in the upper solar atmosphere in comparison with the field present at the photosphere. The LOS magnetic field inferred in the spicules could behave in such a manner, but in this work no clear decrease with height is inferred. The left panel of Fig.~\ref{Figure:16} shows that if the inferred $B_{\mathrm{LOS}}$ from the whole data set \#8 is plotted as a function of the pixel element height, there is no clear correlation between the two values. This has also been found in all other off-limb data sets. It is worth noting that \citet{orozco2015} detected a decrease of the spicules' magnetic field strength from 80~G near the limb to 30~G at 3~Mm height (see their Fig.~4). Above this height there is no clear trend in the magnetic field strength, which is in excellent agreement with the behaviour found here. The right panel of Fig.~\ref{Figure:16} shows the same plot for the chromospheric anemone described in Sect.~\ref{anemonesect}. There seems to be a noticeable decrease of $B_{\mathrm{LOS}}$ as the distance from the limb increases. This decrease amounts to roughly 100~G in 6~Mm, i.e. a $B_{\mathrm{LOS}}$ gradient around 16 G Mm$^{-1}$, which is almost identical to that inferred from the lowest 3~Mm of Fig.~4 of \citet{orozco2015}.

The difference in the behaviour of $B_{\mathrm{LOS}}$ as a function of height for spicules and the anemone could be the length of the former. The signal of spicules comes mostly from heights between 2\arcsec\ and 5\arcsec\ above the limb, while the signal from the anemone comes from heights between 4\arcsec\ and 11\arcsec. The anemone is longer, and while at a height of 4\arcsec\ above the limb both the spicules and the anemone show $B_{\mathrm{LOS}}$ values close to 250 G, at 10\arcsec\ height the values of the latter are close to 150~G. If there were data from spicular material on a wider height range, perhaps the inferred $B_{\mathrm{LOS}}$ would behave in a similar manner. Spicules can have longer lengths than those seen at the \ion{Ca}{II} 8542~\AA\ line, but they become too faint at upper heights in this line.


\begin{table*}
        \centering
        \caption{Results of spatially binning and temporally averaging data set~\#8. For different combinations of pixel scale and integration time, the average $B_{\mathrm{LOS}}$ mode for the whole data set is given together with the percentage of the number of points (with respect to the total number of points available) that pass the two WFA validity criteria.}
        \label{table:dataset8}
        \begin{tabular}{c|c|c|c|c|}
            \cline{2-5}
             & \multicolumn{4}{|c|}{Pixel scale (\arcsec~pixel$^{-1}$)}\\
            \hline
             \multicolumn{1}{|c|}{Int. time} & 0.057 & 0.114 & 0.456 & 0.912 \\
             \hline
            \hline
            \multicolumn{1}{|c|}{36.33 s} & 193 (0.36\%)  & 203 (0.05\%) & 145 (0.18\%) & 117 (0.38\%) \\
            \hline
            \multicolumn{1}{|c|}{5 min} &166 (0.69\%) & 151 (0.81\%) & 165 (0.62\%) & 167 (0.65\%) \\
            \hline
            \multicolumn{1}{|c|}{20 min} &171 (1.70\%) & 170 (1.80\%) & 169 (1.10\%) & 150 (1.31\%) \\
            \hline
            \multicolumn{1}{|c|}{45 min} &175 (1.76\%) & 209 (1.54\%) & 174 (0.82\%) & 180 (0.66\%) \\
            \hline
        \end{tabular}
\end{table*}

\begin{table*}
        \centering
        \caption{Same as Table~\ref{table:dataset8} for data set~\#11.}
        \label{table:dataset11}
        \begin{tabular}{c|c|c|c|c|}
            \cline{2-5}
             & \multicolumn{4}{|c|}{Pixel scale (\arcsec~pixel$^{-1}$)}\\
            \hline
             \multicolumn{1}{|c|}{Int. time} & 0.057 & 0.114 & 0.456 & 0.912 \\
             \hline
            \hline
            \multicolumn{1}{|c|}{36.33 s} & 148 (0.0013\%)  & 142 (0.0002\%) & 104 (0.074\%) & 59 (0.1001\%) \\
            \hline
            \multicolumn{1}{|c|}{5 min} &77 (0.0171\%) & 70 (0.0225\%) & 47 (0.0995\%) & 25 (0.2068\%) \\
            \hline
            \multicolumn{1}{|c|}{20 min} &51 (0.1150\%) & 52 (0.1325\%) & 28 (0.2068\%) & 25 (0.3222\%) \\
            \hline
            \multicolumn{1}{|c|}{45 min} &36 (0.2624\%) & 45 (0.3077\%) & 36 (0.2612\%) & 15 (0.2089\%) \\
            \hline
        \end{tabular}
\end{table*}

\subsection{Integration time and pixel scale} \label{sec:cadence}

As discussed in the Introduction, there are several studies that have tried to infer the magnetic field present in spicular material. However, none of such studies uses data with the same exposure time (or integration time in the case of temporal averaging) and pixel scale as the one presented here. \cite{Centeno2010}, for example, used data with an effective integration time of 45 minutes and 1\arcsec\ pixel scale. \cite{orozco2015} had a short integration time of 10~s and a spatial resolution of the order of 0.7--1\arcsec. Since spicules have a lifetime of a few minutes at most and are thin structures, it is difficult to properly study them with such cadence and pixel scale.

In order to compare how the results of this study would change if the data used were of lower cadence and larger pixel scale, we perform a spatial binning and temporal integration of two of our data sets (\#8 and~\#11). The Bayesian inversion is applied to all pixels of the new data set and then the mode of the $B_{\mathrm{LOS}}$ distribution together with the percentage of points that meet the correlation and asymmetry restrictions are presented in Tables~\ref{table:dataset8} and \ref{table:dataset11} for data sets~\#8 and~\#11, respectively. The top left cell in both tables corresponds to the results presented in Sects.~\ref{sec:off-limb} and~\ref{sec:on-disk}. In the case of off-limb pixels in the vicinity of an active region (Table~\ref{table:dataset8}), the value of \BLOS\ for the shortest integration time decreases as the pixel scale is increased. Nevertheless, this behaviour is not found for the other integration times considered and, in general, the \BLOS\ in Table~\ref{table:dataset8} do not depart too much from those computed with the best cadence and pixel scale. It is interesting to note that when the pixel scale is held fixed, then the more exposures are added, the larger the percentage of pixels that can be used in the Bayesian inversion. This clear trend is not observed when the integration time is fixed and the spatial binning is varied. Regarding on-disk pixels away from an active region (Table~\ref{table:dataset11}), we find that both the spatial binning and the temporal integration yield smaller values of \BLOS. For example, \BLOS\ decreases by a factor 2.5 if we move along the first row from the best to the worst pixel scale and by a factor 4.1 if we move along the first column. Since the percentage of useful pixels increases in both cases, this result looks very solid. A similar behaviour of \BLOS\ with respect to variations of the integration time and pixel scale are found for data set~\#1 (off-limb \& QS), the change of \BLOS\ along the first row and first column now being 2.30 and 1.54, respectively. It must be mentioned, however, that the statistics of data set~\#1 is much weaker because the percentages are considerably lower than those in Table~\ref{table:dataset11}.


\section{Conclusions} \label{conclusions}

In this paper, the Weak Field Approximation (WFA) has been applied for the first time to chromospheric spicules, both on the disk and above the limb. For this purpose we have used spectropolarimetric observations with the SST/CRISP instrument with an exposure time of either 26 s (one data set) or 36 s (five data sets). Given the limitations of the theory and its range of validity, it is clear that a careful treatment of experimental data sets has to be conducted in order to apply the WFA properly. In this study we have translated such limitations into two restrictive criteria, one for the correlation between $V(\lambda)$ and $\partial I(\lambda)/ \partial \lambda$ and another one for the measured asymmetry of the Stokes $V$ profiles. The imposed criteria result in much less on-disk than off-limb pixels being considered for the calculation of their $B_{\mathrm{LOS}}$, something that was expected given the stronger superposition of spectropolarimetric signals on the disk, where the background chromosphere contributes to the observed Stokes profiles \citep[see][and Sect.~\ref{sec:superposition}]{Centeno2010}. This means that the WFA works significantly better for our off-limb observational data sets.

The Bayesian inversion of the Stokes $I$ and $V$ profiles leads to line-of-sight magnetic field components well in excess of 100~G, often reaching as high as 500 G and more. In general, the largest values of $B_{\mathrm{LOS}}$ are found in off-limb spicules, probably as a consequence of the much smaller signal superposition, that is stronger on the disk and therefore reduces the \BLOS\ obtained there. Following \cite{Centeno2010}, we note that the values presented here are a lower limit of the total magnetic field intensity, for three reasons. The first one is quite obvious: having measured just the LOS magnetic field component, the $\vec{B}$ vector has a larger magnitude than this measurement. Second, we have assumed a filling factor equal to one, but if a pixel has $f$ different from one, then the reported value of $B_{\mathrm{LOS}}$ increases by a factor $f^{-1}$. Third, the overlapping of emitting off-limb structures or absorbing on-disk structures often reduces the measured $B_{\mathrm{LOS}}$; see Fig.~\ref{fig:superposition}.

The $B_{\mathrm{LOS}}$ values obtained in this work, most of them at heights between 2\arcsec\ and 5\arcsec\ above the limb, are unprecedented and for this reason we have placed them on a firm basis. We have seen that magnetic fields with |$B_{\mathrm{LOS}}$|> 100 G are abundantly present in on-disk and off-limb spicules, both near and away from active regions (see the number of counts in Figs.~\ref{Figure:11} and~\ref{Figure:12}). We have also proved that the on-disk pixels whose $B_{\mathrm{LOS}}$ we have been able to measure are co-spatial with photospheric magnetic field concentrations in the network boundaries and that the magnetic field orientation agrees in the photosphere and in these pixels. This result does not confirm that our on-disk \BLOS\ measurements can be unequivocally associated with spicules, but provides a strong basis for this association at least in a considerable number of pixels.

We have found no evidence of a difference in magnetic field strength between QS spicules and those near an AR, although this conclusion must be confirmed or refuted because we have only studied one data set with off-limb QS spicules. Furthermore, the sign of the LOS component of the magnetic field seems to be strongly correlated to the location of the spicules. Those close to a sunspot show a uniform sign equal to that of the sunspot, while spicules located far from a region with a clearly reigning polarity present both $B_{\mathrm{LOS}}$ signs \citep[see][ for a similar conclusion]{orozco2015}.

Spicules are small scale and very dynamic structures. They appear and fade in a short time and during their existence they can be subject to transverse motions, that disturb their position on the plane-of-the-sky and their Doppler velocity. Furthermore, the emission or absorption of several spicules can overlap along a given LOS. Then, during the exposure time, which in our case is the time needed to acquire $I$, $Q$, $U$ and $V$, the detector records a non-steady signal coming from one or more time-varying structures. This is the reason why long exposure times or the averaging of several exposures, which leads to a long integration time, can corrupt the four Stokes profiles. The same can be said about the spatial binning of data. We have investigated all these effects in Sects.~\ref{sec:superposition} and~\ref{sec:cadence}. In the first one, the two-slab model of \citet{Centeno2010} has been exploited to understand how signal superposition affects the observed Stokes $I$ and $V$ parameters and the determination of \BLOS\ from them. It is found that observing two overlapping spicules above the limb leads to either \BLOS\ values that are lower limits to those of the slabs or to Stokes parameters that do not meet the two criteria we impose prior to the application of the Bayesian inversion based on the WFA. In Sect.~\ref{sec:cadence} we have proved that the pixel scale and integration time of a data set have a strong influence on the inferred \BLOS\ values and that a coarse spatial and/or temporal resolution lead to smaller \BLOS\ results in QS spicules. This explains why previous works failed to detect such large magnetic fields in spicules in spite of using spectropolarimetric observations, as in this paper. In fact, it is actually surprising that, given the short lifetime of spicules, observations with an integration time of a few minutes \citep[as in][]{TrujilloBueno2005} or tens of minutes \citep{2005ESASP.596E..82R,2006ASPC..358..448R,Centeno2010} allow to calculate their magnetic field. We put forward the hypothesis that the strong spicular magnetic field, perhaps combined with their possible recurrence in the same position, are responsible for the success of these efforts. We suggest that the spicule magnetic field remains relatively unchanged with time and that spicules are seen as a consequence of density being accumulated by some process (be it a shock, instability, etc.). It is at this moment that the opacity of on-disk spicules or the source function of off-limb spicules is large enough that the spectral line can be detected and the magnetic field can be measured. After some time, the spicular material will disappear, but will reappear in the same flux tube at a later time, allowing one to measure the field again.

The application of the WFA relies on the assumption that, for the spectral line used here, $B \ll 2650$ G. This constraint may invalidate the results in the high end of $B_{\mathrm{LOS}}$ values, but the bulk of measurements, in the range [50, 300] G, say, have a high credibility. On the other hand, in the limit of magnetic field strengths below some 50~G, the histograms of Figs.~\ref{Figure:11} and~\ref{Figure:12} present a paucity in the number of \BLOS\ inversions. In Sect.~\ref{sec:superposition} we have shown that the application of the WFA to Stokes signals from randomly oriented overlapping structures can be the cause of this effect and that the similarity of the distributions of Fig.~\ref{Figure:12}, from observational data on a QS area, and the middle panel of Fig.~\ref{fig:superposition}, from a Monte Carlo experiment, can be a sign of the impact of signal superposition. The results of Sect.~\ref{sec:superposition} come from a very idealised model and for this reason one cannot rule out that the combination of noise, spectral resolution and/or integration time of our data sets may yield poor quality Stokes~$V$ signals that prevent the Bayesian inversion from being applied when $|\BLOS| \lesssim$~50~G. We thus interpret the distributions of Figs.~\ref{Figure:11} and~\ref{Figure:12} as a sign that the WFA applied to our data is biased toward large \BLOS\ and that smaller values do exist but it may not be possible to detect them because of signal superposition, noise in the data, etc.

Because of the small range of heights in which the WFA could be applied, we have not been able to assess the possible variation of the spicules' $B_{\mathrm{LOS}}$ with height. Nevertheless, a taller chromospheric anemone present in one of the data sets does show an average gradient of $B_{\mathrm{LOS}}$ of the order of 16 G Mm$^{-1}$. This gradient is very similar to that obtained by \citet{orozco2015} in the lowest 3~Mm of spicules.

The results presented here should be tested by observing spicules in other locations and with other instruments. We anticipate that their confirmation will arrive with the Daniel K. Inouye Telescope in the near future.

\begin{acknowledgements}
MK and RO acknowledge support from the Spanish Ministry of Economy and Competitiveness (MINECO) and FEDER funds through project AYA2017-85465-P. They are also grateful for the travel support received from the International Space Science Institute (Bern, Switzerland) as well as for discussions with members of the ISSI team on ``Observed multi-scale variability of coronal loops as a probe of coronal heating'', led by C.~Froment and P.~Antolin. PA acknowledges funding from his STFC Ernest Rutherford Fellowship (No. ST/R004285/1). This research has made use of SunPy v1.1, an open-source and free community-developed solar data analysis Python package \citep{SunPy2020}. The Swedish 1-m Solar Telescope is operated on the island of La Palma by the Institute for Solar Physics of Stockholm University in the Spanish Observatorio del Roque de los Muchachos of the Instituto de Astrof\'\i sica de Canarias. The Institute for Solar Physics is supported by a grant for research infrastructures of national importance from the Swedish Research Council (registration number 2017-00625). D.K. has received funding from the S\^{e}r Cymru II scheme, part-funded by the European Regional Development Fund through the Welsh Government and from the Georgian Shota Rustaveli National Science Foundation project FR17 323. AAR acknowledges support from the Spanish Ministry of Economy and Competitiveness (MINECO) and FEDER funds through project PGC2018-102108-B-I00.
\end{acknowledgements}


\end{document}